\documentclass[usenatbib]{mn3e} 
\usepackage{amsmath,amssymb}
\usepackage{rotating}
\usepackage{pdflscape}
\usepackage{tablefootnote}   
\usepackage{graphicx}         
\usepackage{gensymb}        
\usepackage{color}               
\usepackage{natbib}
\usepackage{times}
\usepackage{fixltx2e} 
\usepackage[flushleft]{threeparttable}

\newcommand{\ramses}{{\sc ramses}}
\newcommand{\ramsestext}{{\sc ramses} }

\newcommand{\cupid}{{\sc cupid} }
\newcommand{\cupidtext}{{\sc cupid}}

\newcommand{\newhorizon}{{\sc NewHorizon} }

\newcommand{\Msol}{\,{\rm M}_\odot} 
 
\newcommand{\Zsol}{\,{\rm Z}_\odot}

\newcommand{\kpc} {{\,\rm kpc}}
\newcommand{\pc} {{\,\rm pc}} 
 
\newcommand{\cc}{{\,\rm {cm^{-3}}}}

\renewcommand{\vec}[1]{{\mathbf #1}} 
\newcommand{\kmsec}{{\,\rm {km\,s^{-1}} }}
\newcommand{\sfr}{$\langle{\rm SFR}\rangle$ }
\newcommand{\Sfr}{$\langle{\rm SFR}\rangle$}
\newcommand{\sigv}{\sigma_{\rm V}}
\newcommand{\Msi}{M_{\rm \star,i}}
\newcommand{\Mgc}{M_{\rm g,c}}
\newcommand{\dxc}{\Delta x_{\rm c}}

\def\Gyr{\,{\rm Gyr}}
\def\Myr{\,{\rm Myr}}
\def\yr{\,{\rm yr}}
\newcommand{\effsf}{\epsilon_{\rm ff,SF}}
 
\newcommand{\eff}{\epsilon_{\rm ff}}

\newcommand{\tff}{t_{\rm ff}}

\newcommand{\effmed}{\langle\epsilon_{\rm ff}\rangle_{\rm med}}

\newcommand{\sigeff}{\sigma_{\log{\eff}}}

\newcommand{\tsy}{t_{\rm\star,y}}

\newcommand{\msy}{M_{\rm\star,y}}

\newcommand{\mtot}{M_{\rm tot}}

\newcommand{\mgmc}{M_{\rm GMC}}
\newcommand{\rgmc}{R_{\rm GMC}}
\newcommand{\siggmc}{\sigma_{\rm GMC}}
\newcommand{\Siggmc}{\Sigma_{\rm GMC}}
\newcommand{\rhogmc}{\rho_{\rm GMC}}
\newcommand{\ap}{\alpha_{\rm vir}}

\newcommand{\fb}{${\rm SNe+RP+SW}$}
\newcommand{\fbi}{${\rm SNe+RP}$}
\newcommand{\fbii}{${\rm SNe+SW}$}
\newcommand{\fbiii}{${\rm SNe}$}
\newcommand{\fbiv}{${\rm RP+SW}$}
\newcommand{\fbv}{${\rm none}$}
\newcommand{\fbvi}{${\rm \eff=100\%}$}

\newcommand{\sfepmed}{\langle\epsilon_{\rm ff}\rangle_{\rm med}}

\renewcommand{\vec}[1]{{\mathbf #1}}

\title[Location of SF \& GMCs] {Physical Properties and Scaling Relations of Molecular Clouds: the Impact of Star Formation}\author[Kearn Grisdale] {\parbox[t]{\textwidth}{Kearn Grisdale$^1$\thanks{kearn.grisdale@physics.ox.ac.uk} }\vspace*{6pt}\\
  	$^1$ Sub-department of Astrophysics, University of Oxford, Keble Road, Oxford OX1 3RH}
\date{\today}
\begin{document}
\maketitle
\graphicspath{ {Figures/} }
\begin{abstract}
Using hydrodynamical simulations of a Milky Way-like galaxy, reaching 4.6 pc resolution, we study how the choice of star formation criteria impacts both galactic and Giant Molecular Clouds (GMC) scales. 
We find that using a turbulent, self-gravitating star formation criteria leads to an increase in the fraction of gas with densities between $10$ and $10^{4}\cc$ when compared with a simulation using a molecular star formation method, despite both having nearly identical gaseous and stellar morphologies. Furthermore, we find that the site of star formation is effected with the the former tending to only produce stars in regions of very high density ($>10^{4}\cc$) gas while the latter forms stars along the entire length of its spiral arms. The properties of GMCs are impacted by the choice of star formation criteria with the former method producing larger clouds. Despite the differences we find that the relationships between clouds properties, such as the Larson relations, remain unaffected. Finally, the scatter in the measured star formation efficiency per free-fall time of GMCs remains present with both methods and is thus set by other factors.
\end{abstract}

\begin{keywords}
galaxies:evolution -- ISM: clouds -- galaxies:ISM -- galaxies: star formation --galaxies:structure 
\end{keywords}

%=========================================================================================================%
%--------------------------------------------------------------------------- Start of Main Text --------------------------------------------------------------------------------------
%=========================================================================================================%
%-------------------------------------------------------------------------------------------------------------------------------------------------------------------------------------------
%--------------------------------------------------------------------------- Section: Introduction -----------------------------------------------------------------------------------
\section{Introduction}
\label{sect:intro}

Observations over the past four decades have been instrumental in quantifying the properties of Giant Molecular Clouds (GMCs) both in the Milky Way (MW) and other galaxies \citep[e.g.][]{Larson:1981aa,Solomon:1987aa,Heyer:2009aa,Roman-Duval:2010aa,Rice:2016aa,Miville-Deschenes:2017aa}, and clumpy galaxies at high redshift \citep[e.g.][]{Swinbank:2015aa}. These observations have shown that a given population of GMCs have a wide range of properties. For example,  \cite{Heyer:2009aa} and \cite{Miville-Deschenes:2017aa} found clouds masses ($\mgmc$) between  $10$ \& $10^{7}\Msol$, radii in the range $0.5\lesssim \rgmc\lesssim 200\pc$ and velocity dispersions $\siggmc\lesssim14\kmsec$ for clouds in the Galactic disc. Furthermore, relations between the properties of clouds have been found \citep{Larson:1981aa} e.g. $\siggmc\propto\rgmc^{a},\, \mgmc\propto\rgmc^{b}$ and $\siggmc\propto(\rgmc\Siggmc)^{c}$ where $a, b$ and $c$ are constant and $\Siggmc$ is the surface density of a cloud. These relations have since become known as the ``Larson relations'' and have been measured by multiple studies since. 

GMCs are the birth place of stars \citep{Myers:1986aa,Shu:1987aa,Scoville:1989aa,McKee:2007aa} and like their other properties, the efficiency (per free-fall time, $\eff$) that they convert gas into stars varies between clouds. Measurements of the dispersion in $\eff$, calculated from observations, range from $0.3$ and $1\,{\rm dex}$  \citep[depending on the method used, see][]{Lada:2013aa,Evans:2014aa,Salim:2015aa,Heyer:2016aa,Usero:2015aa,Gallagher:2018aa,Leroy:2017aa,Utomo:2018aa,Sharda:2018aa,Lee:2016aa,Vutisalchavakul:2016aa}. Given the diversity in both gas properties and $\eff$ one of the key questions in GMC evolution (and hence galactic evolution) is the interdependency of star formation and the gas making up clouds. As of yet this question has yet to be definitively answered. 

Recent galactic simulations are now able to reproduce the observed distribution of global properties of GMCs in Milky Way like galaxies \citep[see][]{Grisdale:2018aa} however this requires the inclusion of stellar feedback \citep[e.g. stellar winds and supernovae, see ][]{Dekel:1986aa,Efstathiou:2000aa,Hopkins:2014aa,Agertz:2016aa}. How and where stars impact their host clouds depends both on the stellar feedback model used as well as where they form. In the majority of galaxy and cosmological simulations the site of individual star formation and feedback are below the simulation resolution. As a results both are normally approximated by a ``sub-grid'' model, however there are several different numerical methods for implementing both physical processes. 

Observations have found that star formation appears to follow a  ``Schmidt Law''  \citep{Schmidt:1959aa,Kennicutt:1998aa}, i.e. $\dot{\rho}_{\star}\propto\rho^{n}$ where  $\dot{\rho}_{\star}$ is the star formation rate density, $\rho$ is the density of star forming gas and $n$ is a power law index linking $\rho$ to $\dot{\rho}_{\star}$. Blindly applying such a law to an entire simulation would result in star formation throughout the entire simulated volume. This has lead to the necessity to use star formation criteria to limit where in the simulation star formation is permitted to occur. One of the more common and easiest to implement is a requirement that only gas above some critical density is allowed to form stars \citep[][]{Krumholz:2012aa,Renaud:2012aa}. An alternative method is to calculate the amount of molecular gas in a given region, from which to form stars \citep[for example, see][]{Krumholz:2009ab}. More recently there have been several simulations that only forming stars in regions where the gas is self-gravitating or unstable against gravitational collapse \citep[for example see][]{Park:2019aa,Dubois:2020aa}. 

Given the same initial conditions the above criteria can lead to stars forming in different locations or at a different rate. This has the potential to change both galaxy and GMC evolution, i.e. stellar feedback will be injected in different locations and at different rates and thus can have significant impact on the structure of GMCs. 
\cite{Hopkins:2013aa}, henceforth H13, explored the impact of different star formation methods (e.g. density, temperature, molecular, self-gravitating, etc.), finding gas density and morphology depends on the choice of star formation criteria, particularly in simulations modelling galaxy-galaxy interactions. However, they also found that the integrated star formation rates where identical for all star formation models. Their results have implications for future simulations attempting to study both cloud and galactic evolution, in particular it raises the question of: how has the choice of star formation criteria impacted results derived from simulations?

In this work, we explore how the evolution of a galaxy is impacted by the choice of star formation criteria, in particular we compare a molecular gas based model to a turbulent method. We detail the resulting differences and similarities for both galactic scale properties  and the properties of GMCs within a MW-like galaxy. This paper is organised as follows: In Section~\ref{sect:meth} we summarise the simulations used in this work and give an overview of the two star formation methods used throughout. In Section~\ref{sect:results} we outline the properties for both the simulated galaxy as a whole and those of the clouds within. We discuss the implications of our results in Section~\ref{sect:Disc} and present our conclusion in Section~\ref{sect:conc}.

%-------------------------------------------------------------------------------------------------------------------------------------------------------------------------------------------
%--------------------------------------------------------------------------- Section: Method -----------------------------------------------------------------------------------------
\section{Method}
\label{sect:meth}

\subsection{Simulations}
\label{meth:sims}
In this work we simulate two Milk Way-like isolated galaxies using the hydro+N-body, Adaptive Mesh Refinement (AMR) code \ramsestext \citep{Teyssier:2002aa}. The initial conditions of the simulation are identical to the AGORA disc initial conditions \citep[see][]{Kim:2016aa} which include both a stellar and gaseous disc, a bulge as well as a (live) dark matter halo. Cells in the AMR grid are refined when they reach a threshold mass of $\sim9300\Msol$ until they reach the minimum cell size of $\Delta x\sim4.6\pc$. The simulations include models for star formation (see \S\ref{sims:sf}) and stellar feedback (henceforth feedback, see \S\ref{sims:fb}). 

The simulations in this work employ an identical set up to those used in \cite{Grisdale:2017aa,Grisdale:2018aa,Grisdale:2019aa} (henceforth G17, G18 andG19 respectively) to which we refer the reader for full details. Briefly,  the galaxies are simulated in isolation (i.e. neglecting environmental factors) and are embedded within a simulation of volume (of size $L_{\rm box}=600\kpc$). The surrounding corona  has an initial temperature, density and metallicity of $10^{6}{\rm K}$, $10^{-5}\cc$ \& $10^{-2}\Zsol$. The initial metallicity of the galaxy is set to $1.5\Zsol$. Finally, star particle that form in the simulation have an initial mass of $\Msi=300\Msol$. 

It is worth noting that the simulation is initialised with the gas and stellar discs on the edge of gravitational collapse. As a result when the simulation begins the system collapse to find a new equilibrium, which is reached after $150\Myr$. We therefore limit our analysis to $t\geq150\Myr$ (unless otherwise stated).

\subsection{Star Formation}
\label{sims:sf}
We employ two different star formation prescriptions (one for each simulation) which we outline below. 

\subsubsection{Star Formation from Molecular Gas}
\label{sims:kmt}
The first method allows any cell with a molecular hydrogen gas mass, $M_{\rm H_2}\geq \Msi$ to form stars. $M_{\rm H_2}$ is calculated for each computational cell and is given by $f_{\star}f_{\rm H_2}M_{g,c}$, where $f_{\star}=0.9$ is the maximum fraction of a cell's mass that can be converted into a star per simulation time-step,\footnote{It is important to note that the need for $f_{\star}$ is numerical (i.e. to prevent the simulation from producing cells with $M_{g,c}\leq0.0\Msol$) not physical.}  $f_{\rm H_2}$ is the fraction of the gas within a given cell that is molecular and $M_{g,c}$ is the total gas mass of that cell.  $f_{\rm H_2}$ is calculated from the density and temperature of the gas in a cell using the \cite{Agertz:2015ab} implementation (see their \S2.3, equations 2-6) of the KMT09 Model \citep[][]{Krumholz:2008aa,Krumholz:2009ab}. 

Each star forming cell then converts gas into stars following:
%<<<<<----- Equation: SF prescription of simulation ----->>>>>%
\begin{equation}
	\dot{\rho}_{\star}= \effsf f_{\rm H_2}\frac{ \rho_{\rm g,c}}{t_{\rm ff}},
	\label{eq:schmidtH2}	
\end{equation}
%<<<<<----- ---------------------------------------------------- ----->>>>>%
where $\rho_{\rm g,c}$ is the density, $t_{\rm ff}=\sqrt{3\pi/32G\rho_{\rm g,c}}$ is the local free-fall time and $\effsf$ is the local star formation efficiency per free-fall time for the gas within the cell. Eq.~\ref{eq:schmidtH2} is simply a Schmidt law \citep{Schmidt:1959aa} where $\rho_{g,c}$ has been replaced by $f_{\rm H_2} \rho_{\rm g,c}$. For \emph{all} star forming cells $\epsilon_{\rm ff,SF}$ is set to $10\%$. We denote this simulation as MSF.
 
\subsubsection{Star Formation from Turbulent Gas}
\label{sims:turb}

The second star formation prescription employed uses the gas dynamics to determine the rate of star formation in each cell\footnote{To reduce computational expense we require a cell to have a gas density of  $\geq1\cc$ for star formation to occur. At full resolution this corresponds to cell with $M_{g,c}\sim2.4\Msol$ and is therefore more than two order of magnitude below the mass of a star particle.} and is based on the ``Multi-freefall PN Model'' outlined in \cite{Federrath:2012aa} \citep[see also][]{Padoan:2011aa}. Below we briefly outline this model as implemented in this work and refer the reader to  \cite{Federrath:2012aa} for the derivation and discussion of the model.  First, the norm (the gradient of $\vec{V}$ being a tensor) of the gradient of the velocity field, $\sigv$, on the scale $\dxc$ and the \emph{isothermal} sound speed of the cell ($c_{\rm s}$, calculated from gas temperature) are determined. We choose to identify $\sigv$ with this norm for consistency with other studies employing this star formation method. Mathematically, we define $\sigv=\|{\rm d}\vec{V}/{\rm d}\vec{X}\|\Delta x_{c} $, where $\vec{V}$, $\vec{X}$ and $\dxc$ are the velocity field, position vector and size of a given cell respectively. From $\sigv$ and $c_{\rm s}$: 
\begin{equation}
	\alpha=\frac{5}{\pi G \Mgc}\frac{\sigv^2+c_{\rm s}^{2}}{\dxc^2},
	\label{eq:sfvirial}	
\end{equation}
the virial parameter for each cell is calculated, where $G$ is the gravitational constant of the Universe. Combining these quantities, the standard deviation in the gas cell, due to thermal motions with a typical mixture of modes, is found: 
\begin{equation}
	\sigma_{s}=\ln\left(1.0+\frac{0.16\sigv^2}{c_{\rm s}^2}\right).
	\label{eq:sigs}	
\end{equation}
We define the logarithmic density contrast of the gas probability distribution function as $s=\ln{(\rho/\rho_{0})}$ where $\rho_{0}$ is the mean density. From this it is possible to define a critical density above which gas will start to condense into stars,
\begin{equation}
	s_{\rm crit} = \ln\left( \alpha\frac{0.067}{\theta^2}\frac{\sigv^2}{c_{\rm s}^2} \right).
	\label{eq:scirt}	
\end{equation}
Proto-stellar discs produce outflows which reduce the fraction of the surround gas that is able to accrete onto new stars \citep[see][]{Wang:2010aa,Myers:2014aa,Federrath:2015aa,Murray:2018aa}. Such processes are not resolved by our simulations and are therefore modelled using two numerical factors: $\theta$ which accounts for uncertainties in density due to shocks and $\epsilon_{\rm PS}$ which accounts for gas removed by outflows. Throughout this work we use $\epsilon_{\rm PS}=0.5$ \citep[see][]{Murray:2018aa}. 
It is important to note that Eq.~\ref{eq:sigs} assumes the density probability distribution function of the isothermal gas cell is log-normal on scales $<\Delta x_{c}$. 

The star formation efficiency per free-fall time is then computed for each cell using: 
\begin{equation}
	\effsf=\frac{\epsilon_{\rm PS}}{2\phi_{t}}
	\exp\left(\frac{3\sigma_s^2}{8} \right) 
	\left[1 + {\rm erf}\left(\frac{\sigma^2_s-s_{\rm crit}}{\sqrt{2\sigma^2_s}}
	\right)\right],
	\label{eq:sfrff}	
\end{equation}
where ${\rm erf}$ is the error function and $\phi_{t}$ accounts for uncertainties in the timescales within a cell. We adopt updated values (Federrath, private communication) for both $\theta$ and $1/\phi_{t}$ ($0.33$ and $0.57$ respectively) to those given in Table 3 of \cite{Federrath:2012aa}. Finally, using Eq.~\ref{eq:schmidtH2}, with $f_{H_{2}}=1.0$, the star formation rate of a given cell is calculated. 

In summary, this model does not have an explicitly set value for $\effsf$. Instead, for each cell at each time step, the amount of gas which can be converted into stars (per free-fall time) is calculated based on how gravitationally bound the gas within a cell is, via, $\alpha$.  In principle this means for an individual cell $\effsf$ can be \emph{any} value, i.e. $0\leq\effsf\leq \infty$. This star formation criteria has been implemented in several previous studies, e.g. see \cite{Kimm:2017aa,Mitchell:2018aa} and \cite{Trebitsch:2018aa}.
We denote this simulations as TSF. 
%----- Stellar Surface Density Maps -----%
\begin{figure*}
	\begin{center}
		\includegraphics[width=0.77\textwidth]{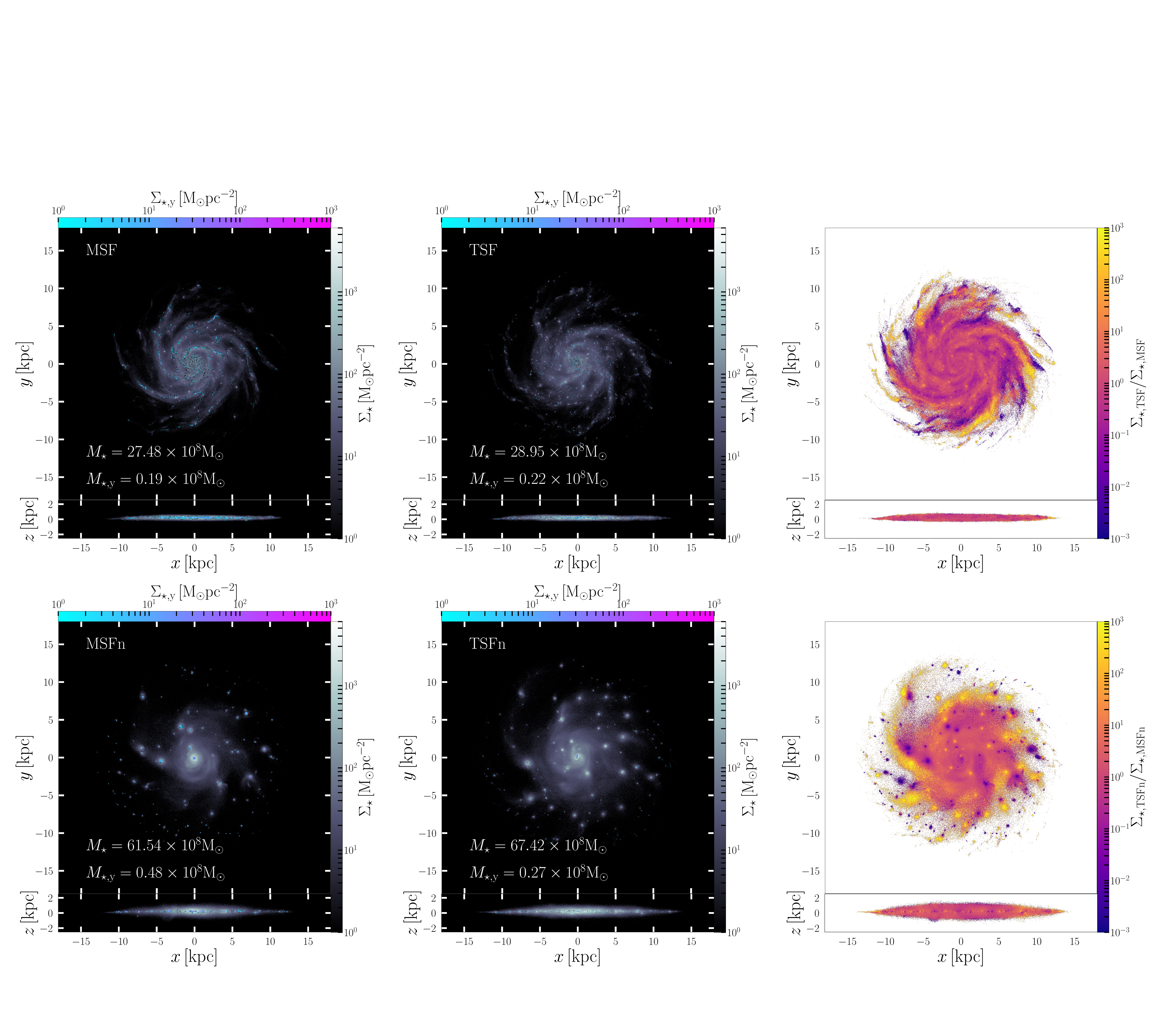}
		\caption{Stellar surface density maps for MSF (left) and TSF (right) at $t=450\Myr$. Each galaxy is shown viewed face-on (top) and edge-on (bottom). The grey scale maps shows surface density calculated using \emph{all} star particles ($\Sigma_{\star}$) while the overlaid surface density map is calculated from star particles that are less than $4\Myr$ old ($\Sigma_{\rm \star,y}$). The total mass of each stellar population is given in each panel. 	
		}	
	\label{fig:sSD}
	\end{center}
\end{figure*}
%----- --------------------------------------- -----%

\subsection{Feedback Model}
\label{sims:fb}
Here we give a brief overview of this feedback model and refer the reader  to \cite{Agertz:2013aa} for the full details. The feedback model assumes that each star particle represented stellar population following a \cite{Chabrier:2003aa}  Initial Mass Function with a total mass of $\Msi$ and accounts for the injection of momentum, energy, mass loss and enrichment from supernovae, radiation pressure and stellar winds from young stars.  

\begin{enumerate}

\item \emph{Supernovae (SNe):}  Both type Ia and II SNe are accounted for (with $\sim15\%$ of all SNe in the simulation being the former). For each type Ia SN that occurs, $0.76\Msol$ of metals, $1.38\Msol$ of gas and $10^{51}\,{\rm erg}$ are injected into the surround environment. Similarly, each type II event injects $10^{51}\,{\rm erg}$, $12\Msol$ of gas with an ejection velocity of $3000\,{\rm km\,s^{-1}}$ and $0.56\Msol$ of metals. The number of type II SNe events per star particle is determined at each time step by calculating the number of ``individual stars'', with mass between $8$ and $40\Msol$, leaving the main sequence. 
Following the SN momentum injection model suggested by \cite{Kim:2015aa} \citep[see also][]{Martizzi:2015aa,Gatto:2015aa,Simpson:2015aa}, if the cooling radius\footnote{the cooling radius scales as $r_{\rm cool}\approx 30 n_0^{-0.43}(Z/Z_\odot+0.01)^{-0.18}$ pc for a supernova explosion with energy $E_{\rm SN}=10^{51}$ erg \citep[e.g.][]{Cioffi:1988aa,Thornton:1998aa,Kim:2015aa}} of an SN explosion is resolved by at least three grid cells ($r_{\rm cool}\geq3\Delta x$), it is initialised in the energy conserving phase by injecting the relevant energy ($10^{51}\,$erg per SN) into the nearest grid cell. If this criterion is not fulfilled, the SN is initialised in its momentum conserving phase, i.e. the total momentum generated during the energy conserving Sedov-Taylor phase is injected into to the 26 cells surrounding a star particle. It can be shown \citep[e.g.][]{Blondin:1998aa,Kim:2015aa} that at this time, the momentum of the expanding shell is approximately $p_{\rm ST}\approx 2.6\times 10^5\,E_{51}^{16/17}n_0^{-2/17} \Msol\kmsec$. 

\item \emph{Radiation Pressure (RP):} The transfer of momentum from photons emitted by stars to the ISM is calculated by assuming that each star particle, (in addition to the stellar population) contains a natal molecular cloud. By modelling the opacity, absorption, scattering of photos, dust heating and how clumpy such a cloud would be, the amount of momentum transferred to the cloud from photons is calculated. Once a particle is older than $3\Myr$ the natal cloud is considered destroyed and momentum contributions from warm dust is no longer calculated. 

\item \emph{Stellar Winds (SW):}  This process models the winds from massive ($>5\Msol$) stars by injecting energy, momentum and gas and metals back into the simulation with the  amount set by approximations to the Geneva high mass loss stellar tracks \citep{Schaller:1992aa,Schaerer:1993aa,Schaerer:1993ab,Charbonnel:1993aa}, see Eq.~4 of \cite{Agertz:2013aa}. Winds from massive stars only contribute to the feedback from star particles with an age less than $6.5\Myr$. SW also includes mass loss from low mass ($\lesssim8\Msol$) stars which provide a source of gas and metals to surrounding environment.
\end{enumerate}

It is worth noting that SW and RP are a continuous feedback mechanics, i.e. they inject gas, momentum and metals into the gas throughout a star particles life. On the other hand SNe provide several individual events which inject gas, momentum and metals into the ISM.

\subsection{Difference From G17-G19}
\label{sims:diff}

The MSF simulations is nearly identical to the simulation ``feedback'' used in G17-G19 however it has two significant changes: the removal of a pressure floor and the choice of $f_{\star}$. In G18 we found that only a small percentage of the gas mass was impacted by the pressure floor (see \S2.1 of G18). However in this work we opt to remove the pressure and to allow just feedback to provide support against gravitational fragmentation. In our previous work $f_{\star}$ was set to $0.5$ which in principle could severely limit star formation in the galaxy. By comparing the results presented in this work with those in G17, G18 and G19 we find that the removal of the pressure floor and the change to $f_{\star}$ have an almost inconsequential impact on the evolution of the galaxy. 

\subsection{Identifying GMCs}
\label{meth:gmcs}

As in G18 and G19, we identify GMCs in projection using the {\sc clumpfind} algorithm \citep{Williams:1994aa} as implemented in the clump finding identification and analysis package \cupidtext\footnote{part of the Starlink Project \citep[see][for details]{Starlink2,Starlink}}. 
Below we briefly outline this method but refer the reader to G17 for the full details and choices of parameters. 

First we create a face-on molecular surface density map of our simulations, where cells with $n\geq\rho_{\rm mol}$ are considered to be purely molecular. The maps are then feed to \cupidtext, which calculated a density contour map from which ``clumps'' (GMCs in this case) are identified. The {\sc clumpfind} algorithm in \cupid has several parameters that determine what can be considered a clump such as: minimum number of pixels, a minimum density contour to be considered and the size of contours. We employ the same \cupid parameters and settings as those used in G18.

Throughout this work, data was obtained from simulation snapshots between $t=150-450\Myr$, separated by $\Delta t=25\Myr$. We employ $\rho_{\rm mol}=100\cc$ and neglect all clouds that lie within the central kiloparsec of the galaxy. 

GMC gas mass ($\mgmc$), radius ($\rgmc$), velocity dispersion ($\siggmc$), young stellar mass ($\msy$) etc, are all calculated using the methods described in G18 and G19, to which we refer the reader. As in our previous work we adopt $\tsy=4\Myr$ as the maximum age of ``young'' stars (unless otherwise stated). Additionally we calculate the virial parameter, 
\begin{equation}
	\ap = \frac{5\siggmc^{2}\rgmc}{G\mgmc},
	\label{eq:virial}
\end{equation}
as well the star formation efficiency per free-fall,
\begin{equation}
	\eff = \frac{\tff}{\tsy}\frac{\msy}{(\mgmc+ \msy)},
	\label{eq:sfeff}
\end{equation}
for each cloud. Here, $G$ is the gravitational constant, $\tff=\sqrt{3\pi/32G\rhogmc}$ is the mean free-fall time across the GMC and $\rhogmc$ is $\mgmc$ divided by the GMC's volume (i.e. its mean density).

%-------------------------------------------------------------------------------------------------------------------------------------------------------------------------------------------
%--------------------------------------------------------------------------- Section: Results -----------------------------------------------------------------------------------------
\section{Results}
\label{sect:results}
%----- Star Formation Histories -----%
\begin{figure}
	\begin{center}
		\includegraphics[width=0.45\textwidth]{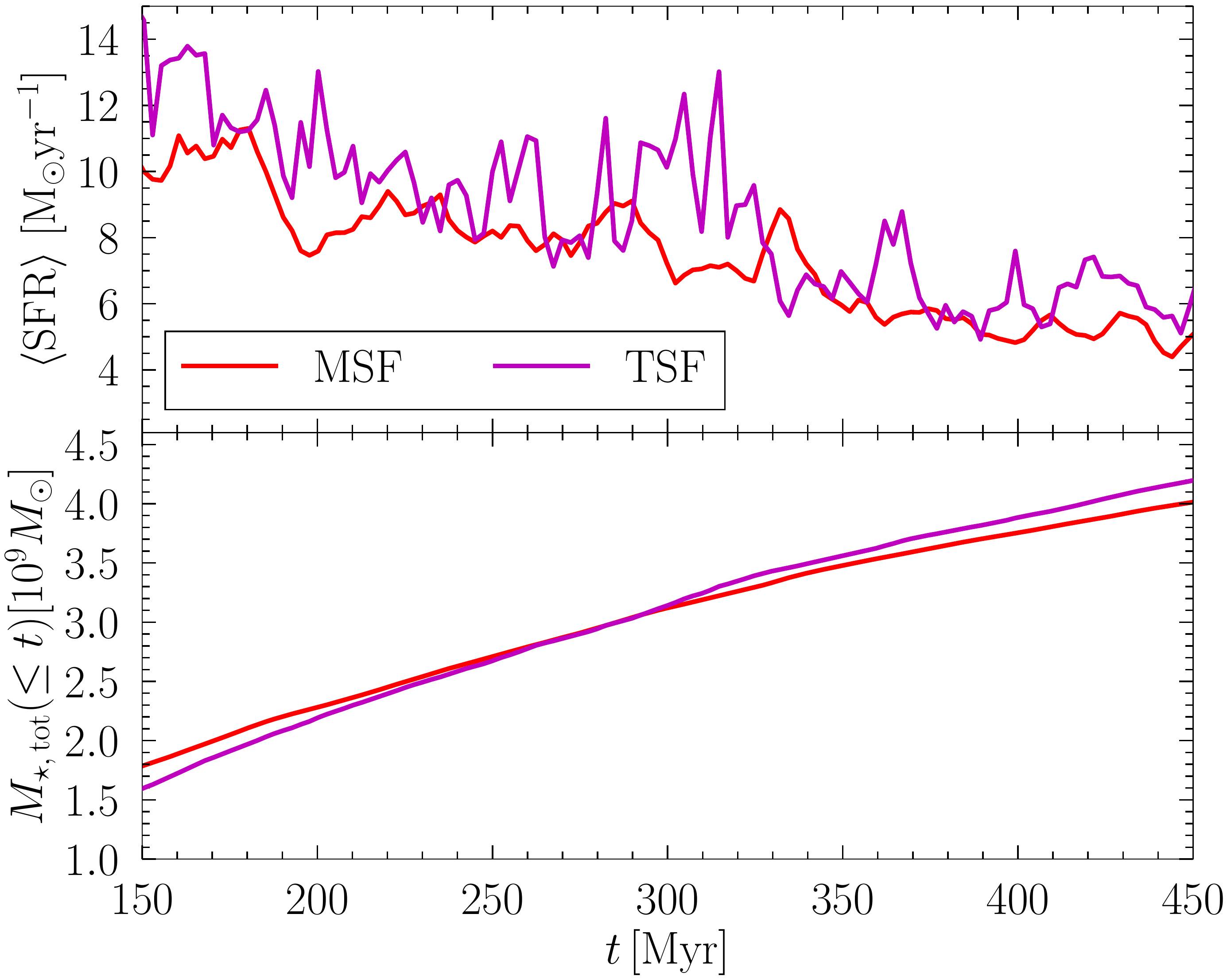}
		\caption{Star Formation History of the two simulations. Top: The mean star formation rate over a $2.5\Myr$ period (\Sfr). Bottom: The cumulative stellar mass formed as a function of time ($M_{\rm\star,tot}$). We only show data after the initial collapse of the simulations (see \S\ref{meth:sims}).
		}
		
	\label{fig:sfh}
	\end{center}
\end{figure}
%----- -------------------------------- -----% 
%----- Stellar PDF Histories -----%
\begin{figure*}
	\begin{center}
		\includegraphics[width=0.65\textwidth]{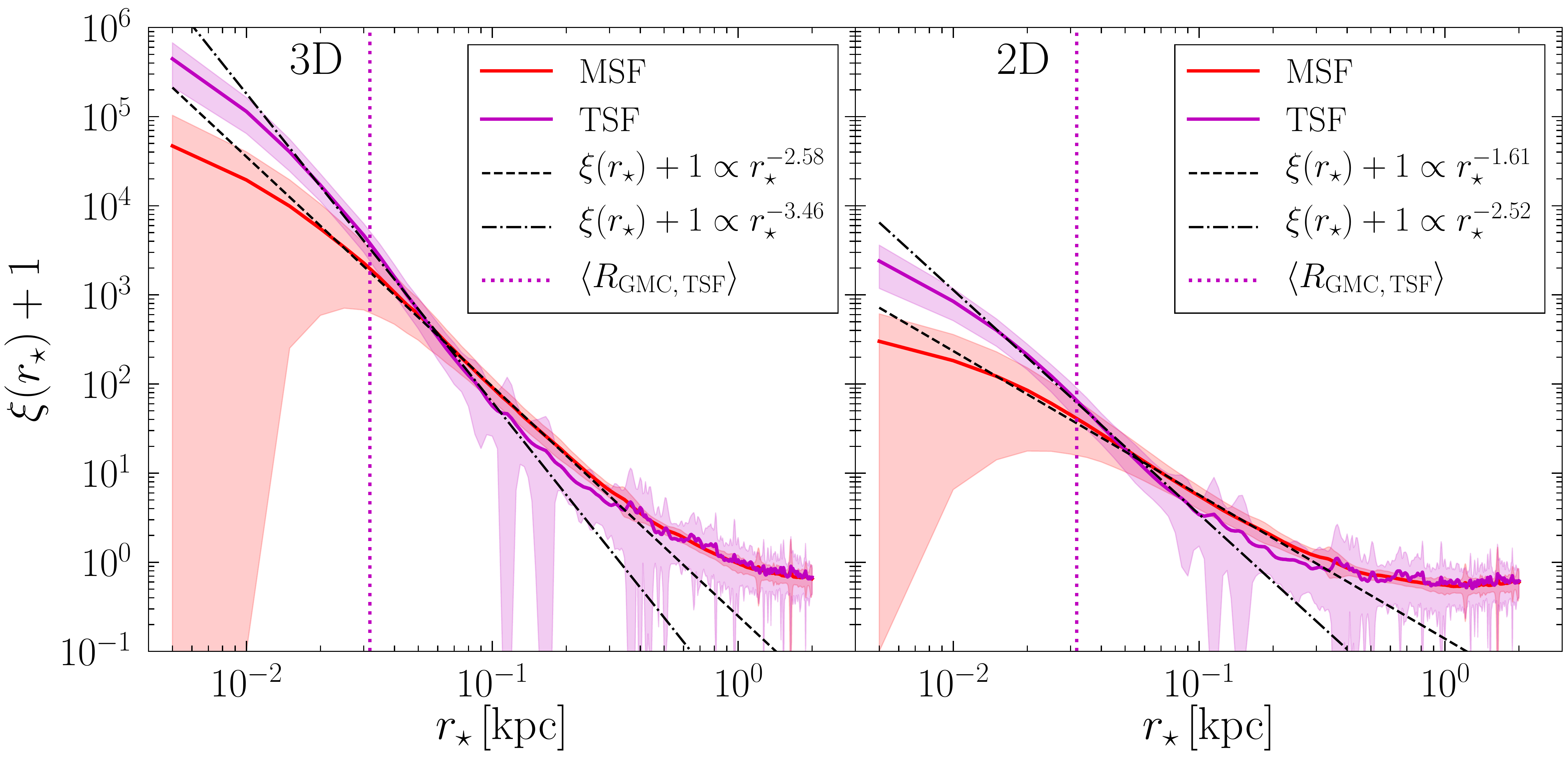}
		\caption{TPCF ($\xi(r_{\star})+1$) for young ($t\leq\tsy$) star particles in the MSF and TSF simulations (red and purple lines respectively). Solid lines give the mean TPCF over a period of $300\Myr$ while the corresponding shaded regions show one standard deviation ($\sigma_{\xi(r_{\star})+1}$) from the mean. In several places the shaded regions cross the lower x-axis, this is a result of $\langle\xi(r_{\star})+1\rangle\leq\sigma_{\xi(r_{\star})+1}$.  The dotted vertical magenta line shows the mean radius of the GMCs found in TSF. The black dashed and dot-dashed line show the a fit to the power law section of the TPCF for MSF and TSF respectively. Left: TPCF calculated using 3D-positions of star particles. Right: TPCF calculated assuming all particles are in the plane of the galaxy, i.e. a 2D projection of particle positions.
		}
		
	\label{fig:stpcf}
	\end{center}
\end{figure*}
%----- ---------------------------- -----%

%--------------------------------------------------------------------------- Section: Results: General Properties --------------------------------------------------------------
\subsection{Impact of Star Formation on Galactic Scales}
\label{res:gp}

While the primary focus of this work is to determine the impact of star formation location on GMCs, it is important to first establish how changing the star formation prescription impacts the galaxy as a whole.

\subsubsection{Location and History of Star Formation}
\label{res:LSF}

Fig.~\ref{fig:sSD} shows the face-on and edge-on stellar surface density ($\Sigma_{\star}$) for MSF and TSF and reveals the two simulations to have remarkably similar stellar morphologies. Both simulations produce a stellar disc with radius of $\sim5\kpc$ with $50\%$ of the galaxies stellar mass contain with a radius of $\lesssim3.4\pc$. Furthermore both galaxies have produce spiral arms, which despite having  slightly in the pitch angles and shape, are relatively similar to each other. 

To identify where star formation is occurring, we plot the surface density of star particles with an age $\leq\tsy$ ($\Sigma_{\rm \star,y}$), on top of the $\Sigma_{\rm\star}$ maps. We find that MSF forms stars along almost the entire length of each spiral arm while TSF forms stars in isolated pockets which trace the spiral arms. Young stars are \emph{slightly} more likely to be found at larger distances above or below the galactic disc in MSF with $|Z_{\rm Gal}|\leq0.1\kpc$ containing $\sim94\%$ of the young stellar mass, while the same height contains $\sim98\%$ in TSF. The difference in the vertical distribution of young stars is just visible in the edge-on panels of Fig.~\ref{fig:sSD}. 

The Star Formation Rate (\Sfr) history, i.e.  the mean Star Formation Rate per $2.5\Myr$ as function of time, is given in Fig.~\ref{fig:sfh} for both simulations. In general the \sfr for both evolves in a very similar way: slowly decreasing from $\sim10-14\Msol\yr^{-1}$ to  $\sim5-7\Msol\yr^{-1}$ over a $300\Myr$ period. The \sfr for TSF, on average, tends to be $\sim1.18\times$ that of MSF, however because of the more bursty nature of the former this only results in $\sim1.05\times$ the total stellar mass by $t=450\Myr$ (values on on Fig.~\ref{fig:sSD}). It is worth noting that while MSF appears to be quiescent compared to TSF, it does have periodic increase in \sfr every $\sim25\Myr$ followed by a subsequent decrease, as discussed below this has impact on the on the gas structure.

As mentioned above, despite the differences in \sfr the two simulations have very similar masses at $t=450\Myr$. The lower panel of Fig.~\ref{fig:sfh} shows cumulative total stellar mass \emph{formed} ($M_{\star,{\rm tot}}$) by each simulation by a given time $t$. Here ``formed'' indicates we are using $\Msi$ of each star particle not their live mass (i.e. their mass after feedback). Both TSF and MSF have a fairly smooth stellar mass growth, with the former having a higher growth rate (i.e. $\frac{{\rm d}M_{\rm\star,tot}}{{\rm d}t}$) than the latter ($\sim 6.2$ and $5.4\Msol\yr^{-1}$ respectively). During the period of initial collapse MSF is able to convert more gas into stars than TSF, creating the offset seen at $t=150\Myr$. Finally, we note that $\frac{{\rm d}M_{\rm\star,tot}}{{\rm d}t}$ for both galaxies decreases as the galaxies use up their gas supply. 

%----- Gas Surface Density Maps -----%
\begin{figure*}
	\begin{center}
		\includegraphics[width=1.0\textwidth]{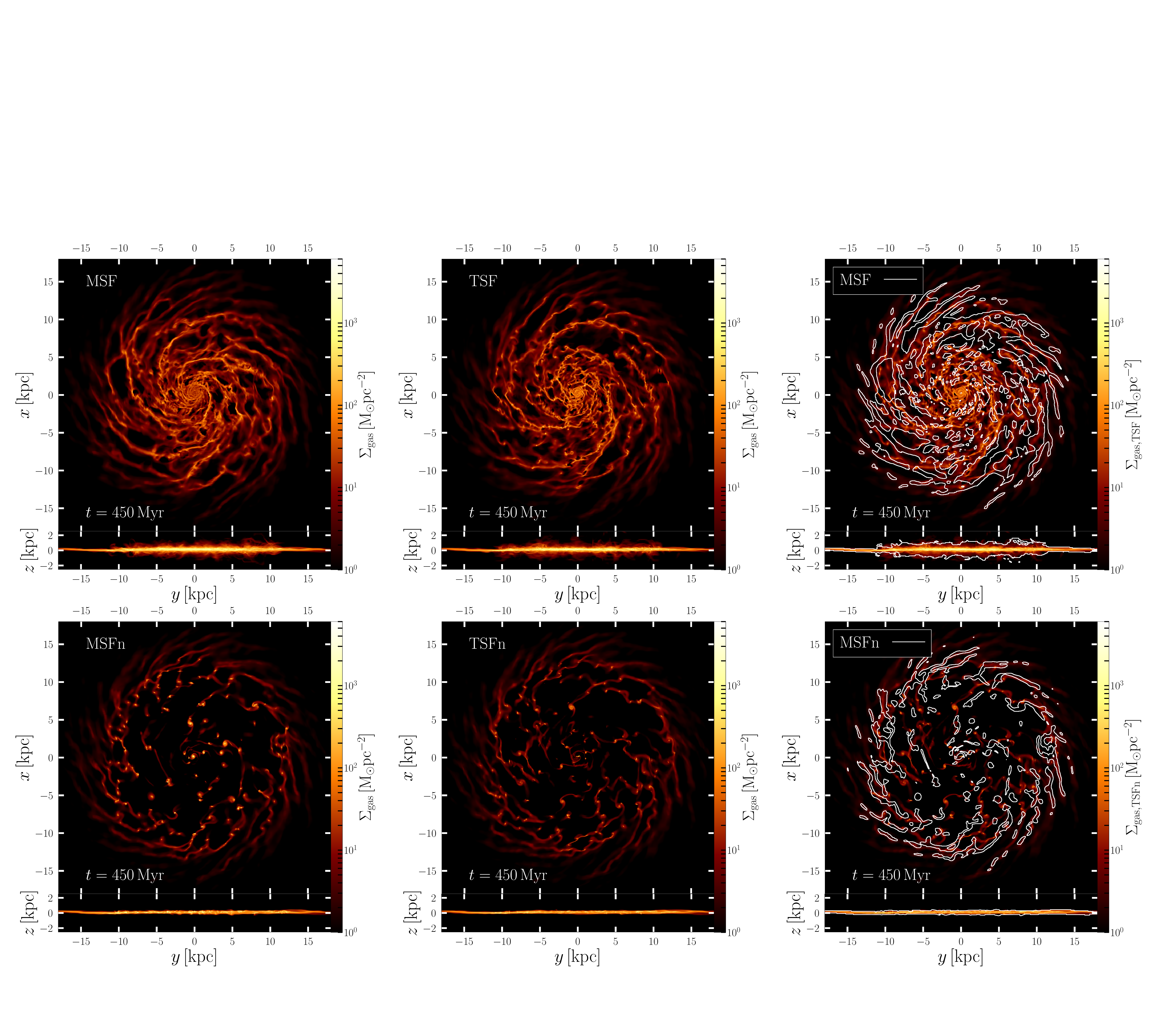}
		\caption{Gas surface density ($\Sigma_{\rm gas}$) maps for MSF (left) andTSF (middle). Each galaxy is shown viewed face-on and edge-on. All panels use the same colour scale and are calculated when the simulation is at $t=450\Myr$. The right hand panel repeats the middle panel but with an added contour representing MSF overlaid. The contour shows gas with $\Sigma_{\rm gas}\geq3\Msol$.
		}
		
	\label{fig:gSD}
	\end{center}
\end{figure*}
%----- ------------------------------------ -----%
The value of $\effsf$ for each cell in TSF is set by its properties and thus can vary greatly (see \S\ref{sims:turb}). For any given $50\Myr$ period we find mean $\effsf$ of $\sim0.075$, a median of $\sim0.06$ and a quartile range spanning from $\sim0.025$ to $\sim0.10$. A handful of cells are able to form stars with an $\effsf$ as low as $\sim10^{-6}$. The largest value of $\effsf$ recorded during run time was $1.95$. 

By calculating the Two Point Correlation Function (TPCF), $\xi(r_{\star})$, of the young star particles we are able to determine if clustering of star formation is different in the two simulations. To estimate $\xi(r)$ we adopt the same definition as \cite{Buck:2019aa}, i.e.
\begin{equation}
	\xi(r_{\star}) = \frac{D_{\rm data}(r_{\star})}{D_{\rm Poisson}(r_{\star})} - 1, 
\end{equation}
where $D_{\rm data}(r_{\star})$ is the number of particle pairs with a separation between $r_{\star}$ and $r_{\star}+\delta r_{\star}$ with $\delta r_{\star}=5\pc$. $D_{\rm Poisson}$ is number of particles pairs, draw from a Poison distribution of particle positions. Fig.~\ref{fig:stpcf} shows $\xi(r_{\star})+1$ when averaged over $300\Myr$. With the exception of the right panel of Fig.~\ref{fig:stpcf} all TPCF presented in this work using the 3D position of each particle to determine $r_{\star}$.

$\xi(r_{\star})+1$ for both simulations follows a reversed Sigmoid function which overlap at $r_{\star}\gtrsim200\pc$. TSF's $\xi(r_{\star})+1$ tends to fluctuate more with time for at $r_{\star}>0.1\kpc$.
By taking the mean size ($\langle R_{\rm GMC,\,TSF}\rangle$) of GMCs identified in TSF\footnote{The mean for MSF and TSF are measured to be $25.8\pc$ and $31.8\pc$ respectively.  We take the larger of these two values to be the upper size limit of an average cloud.} (see \S\ref{res:2D}) it is possible to approximately distinguish between clustering occurring within GMCs (i.e. $r_{\star}\leq\langle R_{\rm GMC,\,TSF}\rangle$) and the clusters of the GMCs with in larger galactic structures (i.e. $r_{\star}> R_{\rm GMC,\,TSF}\rangle$). On the scale of GMCs the TPCF for TSF is $\sim2.1\times$ that of MSF. This suggests that there is more clustering and thus stellar sub-structures within GMCs in TSF. The separation further increase so that on scales comparable to the simulation resolution (i.e. $r_{\star}=5\pc$) the TPCF of TSF is $\sim9.5\times$ large than MSF's.

To further quantify the difference between the two TPCFs we measure the power law index for the power law section of MSF and TSF's TPCFs  (i.e. $0.03\lesssim r_{\star}\lesssim0.2$ and $0.015\lesssim r_{\star}\lesssim0.15$ respectively) and find values of $-2.58$ and $-3.46$ respectively. 
In summary, while both simulation tends to form clusters of star particles, clustered star formation is more common in TSF particularly for $r_{\star}<0.2\kpc$.

\cite{Buck:2019aa} found that the TPCF for their simulations were described by an ``power-law with an exponential cut-off''. They measure power law indexes in the range of $\sim-0.027$ to $-0.094$, which is significantly shallower than those found in our simulations. There are two factors that could affect these measured  power law indexes: age cut off for ``young'' stars and spatial scales coved. In the case of the former they consider stars with ages of $\leq40\Myr$ young \citep[see][]{Wang:2015aa} while we use $\leq4\Myr$. If we recalculate the TPCF using $\tsy=40\Myr$ and find that the power law index for both simulations is reduced by $\sim 1$. We note that their TPCF are calculated for $0.1\leq t_{\star}\leq6\kpc$ compared to our  $0.004\leq t_{\star}\leq3\kpc$. It is unclear if they don't see clustering on smaller scales or if they do how this clustering is dealt with, which makes determining if our results are comparable challenging.  

Perhaps a more important comparison is to the TPCFs presented in \cite{Grasha:2017aa} from observations of several galaxies. In general the shapes of their TPCF's match ours, i.e. flat at large scales ($\gtrsim0.5\kpc$), a power law followed by a possible flattening at small scales ($\lesssim0.01\kpc$). From a visual inspection NGC 3344 appears most similar to our simulations, i.e. it is a relatively face-on galaxy with similar spiral structure (see their Fig.~1). For NGC 3344 they find a power law index of $-1.68$ for $5<r_{\star}\lesssim43\pc$, which is shallower that what we find. This discrepancy is largely explained by the fact that our TPCF's are calculated using the 3D position of particle while the TPCFs in \cite{Grasha:2017aa} are calculated in 2D projection. The right panel of Fig.~\ref{fig:stpcf} shows the TPCF when $r_{\star}$ is calculated neglecting each particles height above the galactic disc, i.e. in projection. When this is done we find that the power law index of the TPCF for both simulations has decreases by $\sim1$ and MSF is now very good match to observational data from \cite{Grasha:2017aa}, suggesting that TSF maybe suffering from over clustering on small scales. That being said, as when comparing with the simulations of \cite{Buck:2019aa}, it is important to note the age of the stars being considered in the TPCF calculations by \cite{Grasha:2017aa}. Their Fig.~3 seems to indicated that stars with ages ranging from $\sim1\Myr$ to $1\Gyr$ are included. As discussed above including older stars leads to shallower TPCFs and thus if calculations for NGC 3344 only included those that are younger than $4\Myr$, i.e. matching our value of $\tsy$, we would see steeper power laws. Due to both the 2D projection and the age of particles it is difficult to draw direct comparisons between the TPCFs measured from our simulations and those in observations, thus we are unable to say if TSF is indeed producing stars that are too spatial clustered.

\subsubsection{Galactic Gas Structures}

In Fig.~\ref{fig:gSD} we show face-on and edge-on gas surface density ($\Sigma_{\rm gas}$) maps for MSF and TSF. A visual comparison of the two galaxies reveals them to also have a very similar gas structures, i.e. they have approximately the same number of spiral arms with similar lengths, positions and pitch angles. Plotting the $\Sigma_{\rm gas}$ map of MSF (as a contour map) on top of TSF's $\Sigma_{\rm gas}$ map (Fig.~\ref{fig:gSD}, right) demonstrates how similar these galaxies are.  Despite the similarity in structure, we find that the maximum $\Sigma_{\rm gas}$ reached by TSF is approximately twice that of MSF. Furthermore we note that the spiral arms of TSF  have the appearance of ``beads on a string'', while MSF has a more even gas distribution. 
There is also a noticeable difference in the size and structure of the galactic centre. Finally we note that the vertical structure of these galaxies are nearly identical, with any dissimilarities easily explained by differences in where stars form and hence where feedback is injected. 
%----- PDFs -----%
\begin{figure}
	\begin{center}
		\includegraphics[width=0.48\textwidth]{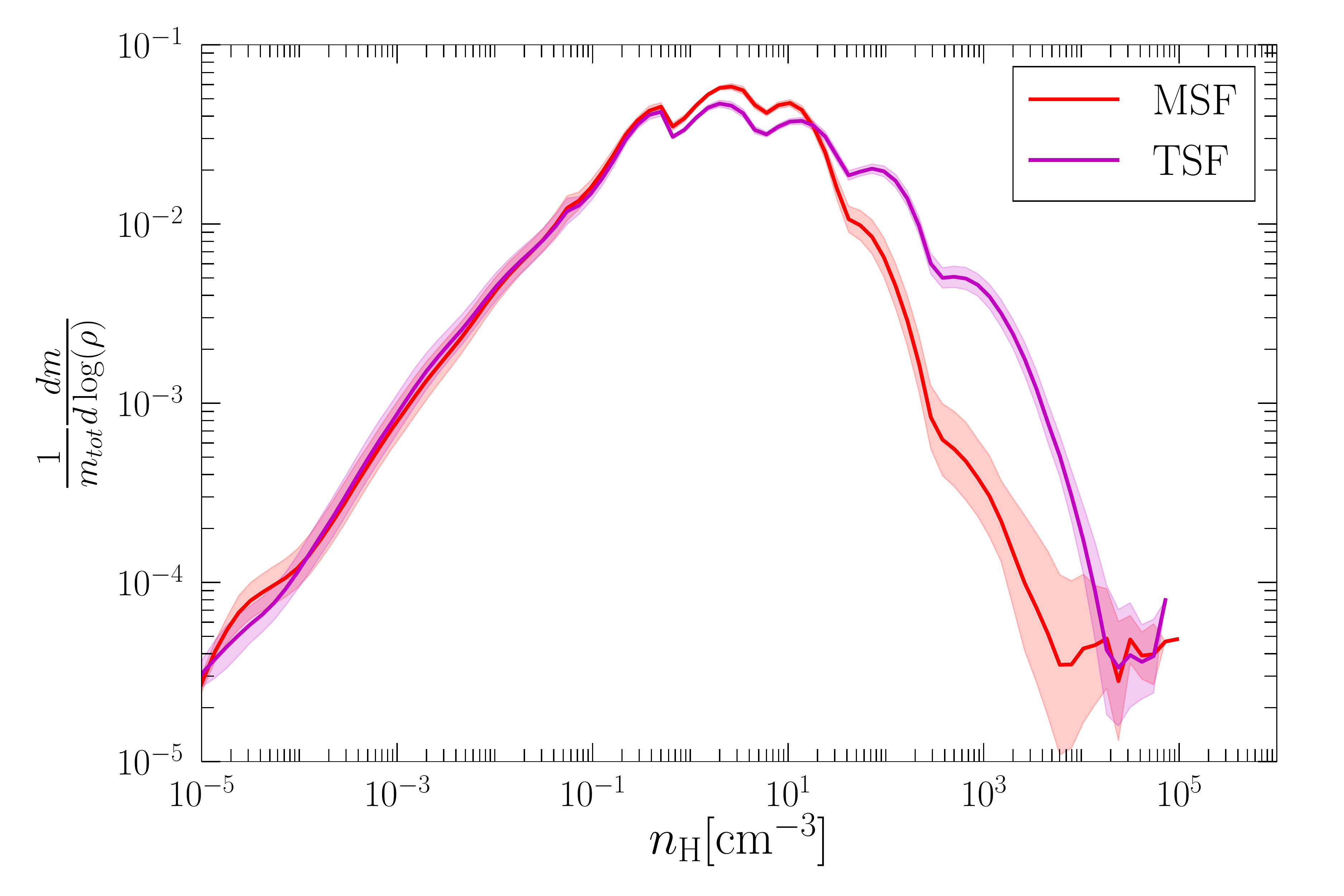}
		\caption{Gas density PDFs of the two simulations. The solid lines show the mean PDF over a period of $300\Myr$. The corresponding shaded regions show one standard deviation from the mean. NB: The regular ``bumpy''  structure seen in both PDFs is a numerical effect resulting from the AMR grid and should be ignored.
		}
		\label{fig:pdf}
	\end{center}
\end{figure}
%----- ------- -----%

By calculating the Probability Distribution Function (PDF) for the gas density, we are able to quantitatively compare the gas structures of these galaxies. In Fig.~\ref{fig:pdf} we show the mean PDF averaged over $300\Myr$ for each simulation. For densities of $n_{\rm H}\lesssim1\cc$ MSF and TSF are nearly identical, while in the spiral arms ($10\lesssim n_{\rm H}\lesssim10^{4}\cc$) TSF has a larger fraction of gas at a given density.  This difference is explained by location of star formation: as described above MSF forms stars along the lengths of its arms and thus depletes the gas throughout the entire arm. TSF, on the other hand, only removes gas in small regions and therefore more gas remains at such densities. 
At the most extreme densities ($n_{\rm H}>10^{4}\cc$) the two simulations tend to have a similar fraction of gas at a give density. 

Fig.~\ref{fig:pdf} also gives an insight into how the density structures evolve: for $n_{\rm H}\lesssim100\cc$ the gas is in a steady state with very little variation, while for denser gas variations are seen on the order of $0.1-1 \,{\rm dex}$. In particular MSF shows considerable variation for $n_{\rm H}\gtrsim10^{3}\cc$ compared with TSF.  
This variation is a direct result of the $\sim25\Myr$ cycles in the \sfr described above: the periodic increases in \sfr leads to an increase in feedback which in turn lead to periodic changes in the PDF. TSF, on the other hand, has \sfr variations on time scales of $2-5\Myr$ which leads to a more continuous and consistent injection of feedback. 

%----- FB Cumalitive Mass -----%
\begin{figure*}
	\begin{center}
		\includegraphics[width=0.8\textwidth]{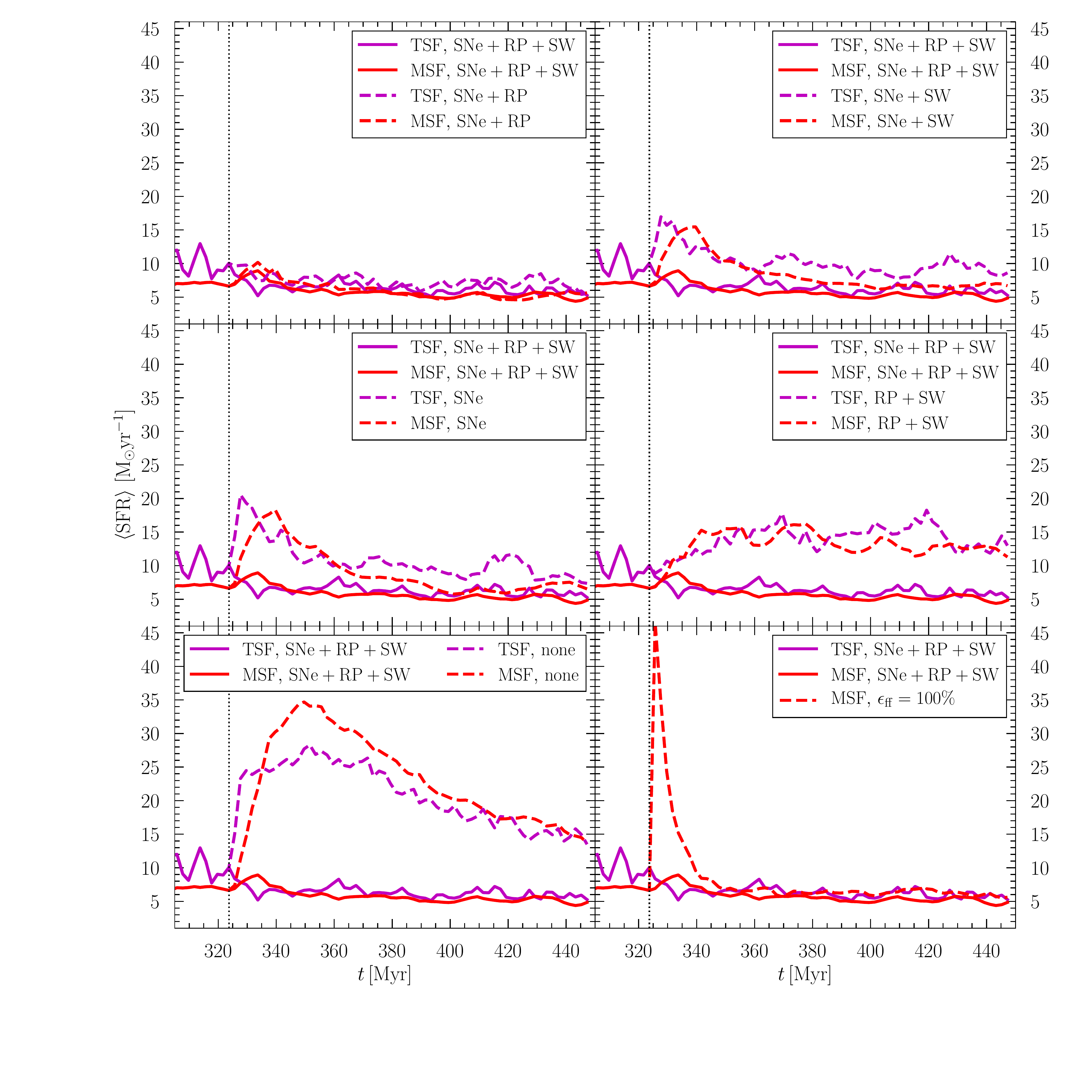}
		\caption{Comparison of the star formation rate (\Sfr) for different feedback models. Each panel compares the fiducial MSF and TSF simulations (solid red and magenta lines) with one other feedback combination (dashed lines). The bottom right panel only compares the fiducial simulations to MSF run with $\eff=100\%$. The vertical dashed lines show the point at which the simulation switched from the fiducial feedback model to the one being tested, i.e. $t=325\Myr$. 
		}
		\label{fig:fbcm}
	\end{center}
\end{figure*}
%----- ------- -----%

%----- Stellar PDF Histories -----%
\begin{figure*}
	\begin{center}
		\includegraphics[width=0.7\textwidth]{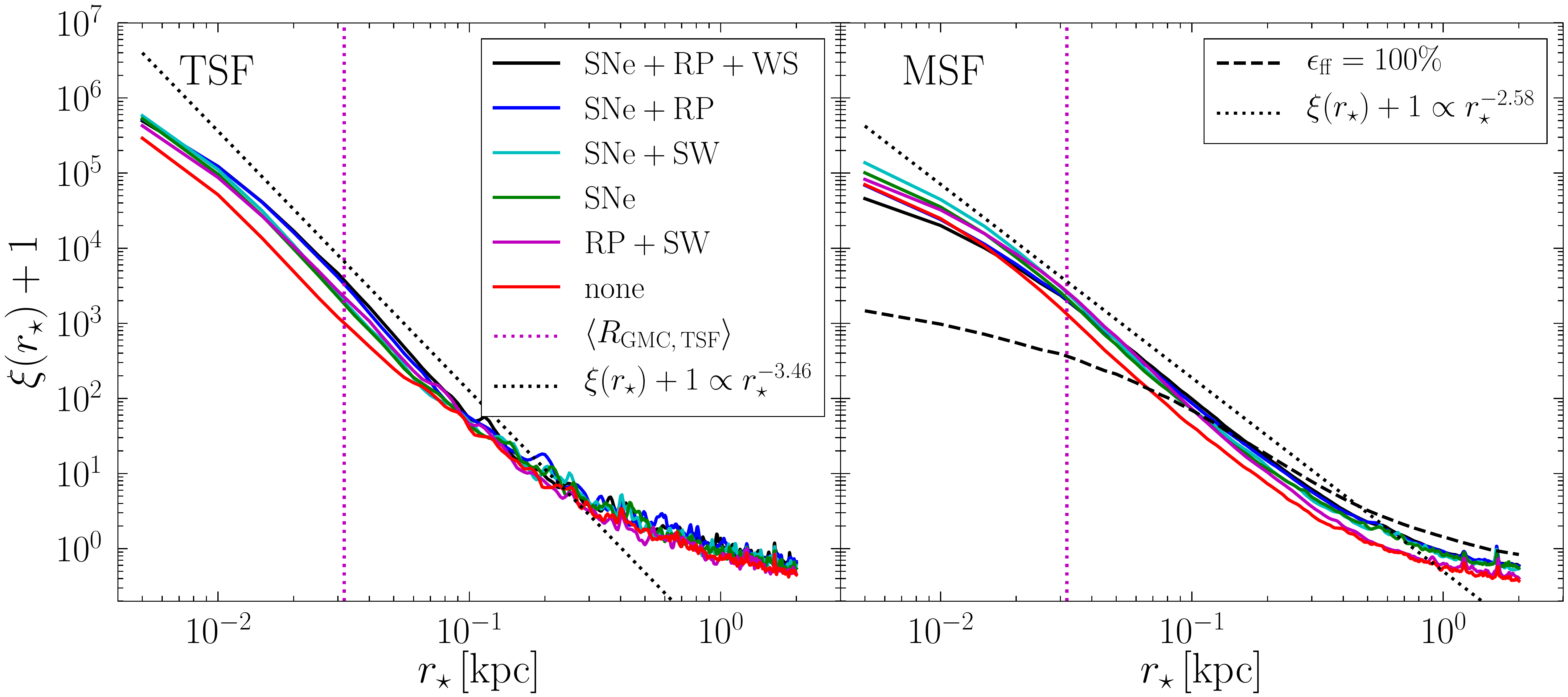} 
		\caption{Comparison of TPCF ($\xi(r_{\star})+1$) for different feedback models. The left panel compares feedback models in the TSF simulations, while the right panel compares those models in the MSF simulations.  Each lines represents the mean TPCF over a period of $130\Myr$. The dotted vertical magenta line shows the mean radius of the GMCs found in TSF. To guide the reader the fits from Fig.~\ref{fig:stpcf} are included as the black dotted lines. 
		}
	\label{fig:stpcffb}
	\end{center}
\end{figure*}
%----- ---------------------------- -----%

%----- FB Cumalitive Mass -----%
\begin{figure*}
	\begin{center}
		\includegraphics[width=1.0\textwidth]{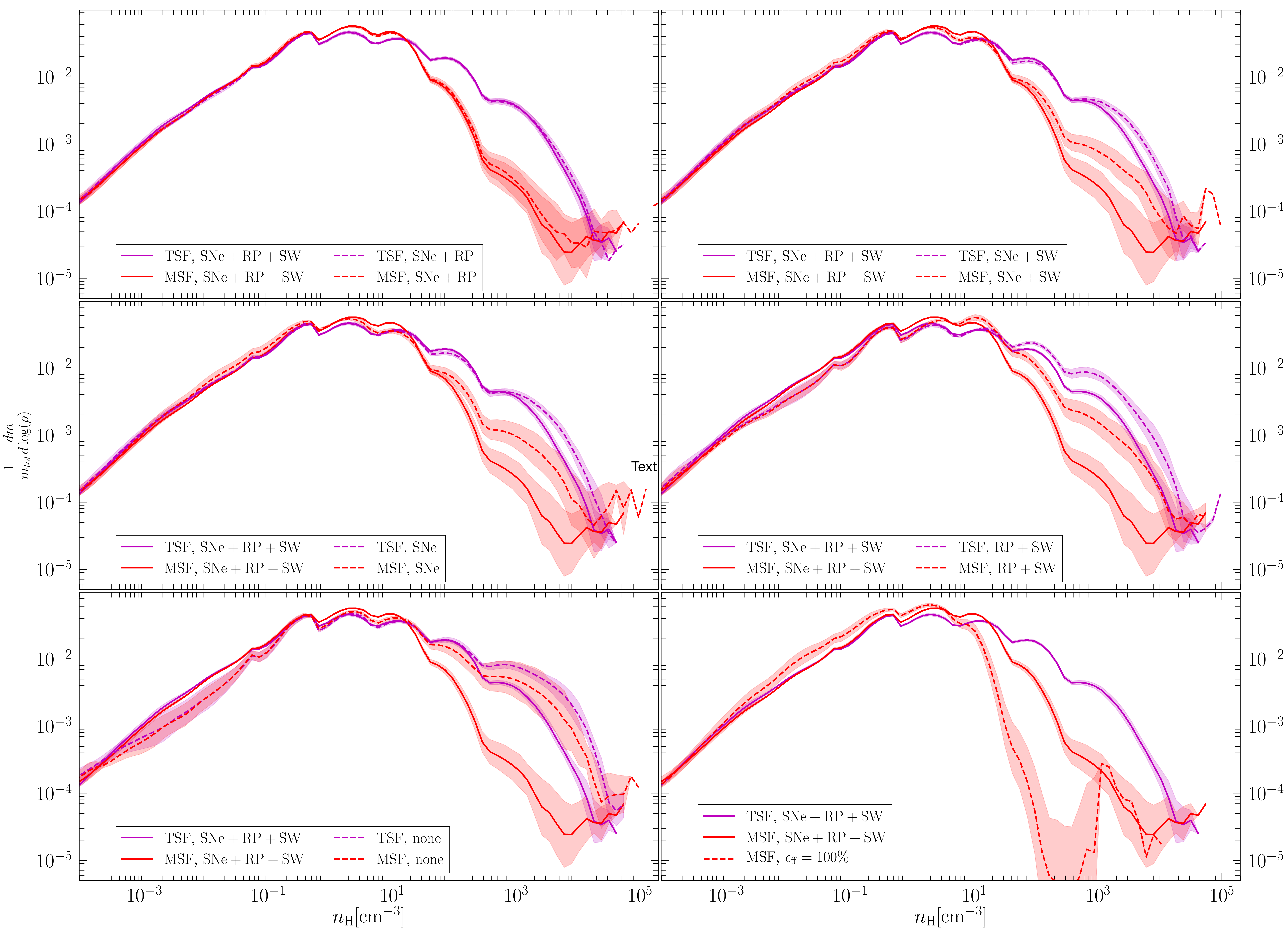}
		\caption{Comparison of the gas density PDF for different feedback models. Each panel compares the fiducial MSF and TSF simulations (solid red and magenta lines) with one other feedback model (dashed lines). The bottom right panel only compares the fiducial simulations to MSF run with $\eff=100\%$. Lines represent the average PDF for $325\lesssim t \lesssim 450\Myr$ and the shaded regions represent one standard deviation from the mean.
		}
		\label{fig:fbgpdf}
	\end{center}
\end{figure*}
%----- ------- -----%

\subsection{Star Formation or Feedback?}
\label{res:fb}

We now explore the role that feedback from star particles plays in changing the galactic structure as described above. 
\subsubsection{The Impact of Feedback on Star Formation and Stellar Structure}
\label{fb:sfh}
As outlined in \S\ref{sims:fb}, the feedback model used in the work consists of several components, each of which can be turned off individually. By rerunning the simulation from $t=325\Myr$ until $t=450\Myr$ with different combination of feedback components (which we will refer to as different models) we are able to determine how the difference in stellar TPCF, \Sfr, morphologies and gas PDF are linked to feedback. To this end we ran an additional five different combinations of feedback: 
\begin{enumerate}
	\item \fbi,
	\item \fbii,
	\item \fbiii,
	\item \fbiv, 
	\item \fbv{} (i.e. no feedback),
\end{enumerate}
 in addition to our fiducial run (i.e. \fb) for both TSF and MSF.  The panels of Fig.~\ref{fig:fbcm} compares one of the above models with the fiducial runs of MSF and TSF.  In general moving from \fb sequentially through to \fbiv{}  is akin to reducing the strength and impact of feedback on the galaxy, as shown by the increase in \sfr at $t=450\Myr$, which in turn leads to an increase in $M_{\star,{\rm tot}}(t=450\Myr)$, when moving through the panels of Fig.~\ref{fig:fbcm} in the same order. The first four new feedback models produce a similar offset between offset between the $M_{\star,{\rm tot}}(\leq t)$ of TSF and MSF as that seen in the fiducial models. 

Removing all feedback leads both simulations to experience a sudden growth in star formation as the support against gravitational collapse is removed. This provides the only case where MSF begins to produce more stars than TSF after $t\sim300\Myr$. For $t\gtrsim350\Myr$ both simulations show a decrease in \sfr which is a direct result of the galaxies burning through their gas supply. In addition to switching off feedback at $t=325\Myr$ we have run a version of TSF with no feedback and two versions of MSF without feedback from $t=0$, one with $\effsf=1\%$ and the other with $\effsf=10\%$ (see \S\ref{sims:kmt}). The former choice of $\effsf$ matches the observationally measured value \citep[][]{Kennicutt:1998aa, Krumholz:2007aa,Bigiel:2008aa}, while the latter value matches our fiducial MSF simulation. Comparing these runs without feedback shows the same result as shown in the bottom left panel of Fig.~\ref{fig:fbcm}: MSF produces stars at a much faster rater and quickly burns through its gas supply. In all cases, this is a result of the collapsing gas with sufficient density to form stars (thus meeting MSF star formation requirements) but having sufficient turbulence to not be considered self gravitating and therefore not meeting the TSF star formation criteria.

As an additional check, we rerun MSF from $t=325\Myr$ with $\effsf=100\%$ and all three feedback components. In principle this allows MSF to form stars more quickly and hence have an increase in amount of feedback. As expected there is a quick rise in \sfr before settling to a $\sim6\Msol\yr^{-1}$, which is only $\sim1.2\times$ that of the fiducial MSF. This indicates that the feedback model used in this work will quickly regulate the star formation to a \sfr$\sim5-6\Msol\yr^{-1}$ for any given value of $\effsf\geq10\%$.

Fig.~\ref{fig:stpcffb} gives the time averaged TPCF for the different feedback models. For TSF the shape of the TPCF remains relatively constant between different feedback models, indeed all feedback models seem reasonably consistent with the power law measured for the fiducial feedback run. Here feedback seems to regulate the magnitude of clustering, i.e. at $r_{\star}=\langle R_{\rm GMC,\,TSF}\rangle$ we find the the TPCF for \fb{} is $\sim2.1\times$ that with \fbiii{} only.
 This is not the case for MSF, and in-particular the simulations without feedback and \fbvi. Without feedback, the clustering of star formation is increased, the opposite of what occurs for TSF! This is simply a result of more cells having sufficient mass to form stars (see \S\ref{fb:gas} for more details). In G17 we found that $\gtrsim66\%$ of the turbulent gas motion in simulations without feedback was in the solenoidal mode (see Fig.~13 of G13), while feedback reduced this to $<60\%$. This increase in solenoidal motions provides support against gravity (i.e. $\sigma_{\rm V}$ is larger in Eq.~\ref{eq:sfvirial}--\ref{eq:scirt}) and thus less stars form in a given region (see \S\ref{sims:turb}). The \fbvi{} run of the MSF, leads to a significant decrease (the \fbvi{} simulation has TPCF that is $0.17\times$ that of \fb {} at $r_{\star}=\langle R_{\rm GMC,\,TSF}\rangle$) in clustering of young star particles. This is a direct result of more cells having sufficient mass to form a star particle: i.e. with $\effsf=10\%$ a computation cell must have a minimum of $\sim3333\Msol$ of molecular gas to form a particle but this drops to just $\sim333\Msol$ with \fbvi.

From the evolution of \sfr and TPCFs with different feedback models we conclude that the difference in the star formation history and thus the total stellar mass is a result of how stars are formed in the simulation. However, feedback does a play role in the amplification or suppression of the differences in the star formation histories of these two simulations. 

\subsubsection{The Impact of Feedback on Gas Structure}
\label{fb:gas}
Using PDFs we are able to assess how the gas structure of the galaxy is likewise impacted by the changing the feedback model, see Fig.~\ref{fig:fbgpdf}. For $n_{\rm H}\lesssim1\cc$we again find that MSF and TSF are nearly identical to each other when using the same combination of feedback. Not only is this true for the time averaged mean PDF but also for the size of the temporal variations (i.e. the standard deviation from the mean PDF), as shown by the shaded regions. For $1\lesssim n_{\rm H}\lesssim3000\cc$ with any of the feedback models both MSF and TSF tend to have a higher fraction of gas at a given density than their fiducial run. As a result there is always a  separation between an MSF PDF and its TSF counter part. When feedback is removed altogether this separation is massively reduce but still remains.

At very high densities ($n_{\rm}>10^{4}\cc$) the PDF for all of the non-fiducial feedback combinations suddenly turns up, and in most causes MSF over shots TSF. 
With weaker feedback there is less support against gravity and more gas is able to reach higher densities. Due to its fixed value of $\effsf$, MSF allows gas to build up at high densities before being converted into stars, with weaker feedback this effect is amplified, producing the up turn at very high densities. The $\effsf=100\%$ run for MSF further serves to highlight this: the maximum density reached in all but one snapshot is $~10^{3}\cc$ (at $t=325\Myr$ the simulations is still adjusting to the change in feedback and is able to reach $n_{\rm H}\sim10^{4}\cc$) rather than the fiducial runs maximum of $5\times10^{4}\cc$. TSF does not have the fixed$\effsf$ and is thus able to convert a much large fraction of gas per free-fall time into stars, resulting in a smaller up tick at high densities.  
In addition to the reduced maximum density the MSF run with \fbvi{} shows a markedly different PDF than our fiducial MSF simulation: the fraction of gas found at $3\lesssim n_{\rm H}\lesssim10^{3}\cc$ is greatly depleted. For example, the fraction of gas at $n_{\rm H}=200\cc$ for the run with \fbvi is more than two dex smaller than the fiducial run. 

In summary comparing the gas density PDFs of our MSF and TSF simulations run using different feedback models shows that distinction between TSF and MSF remains for $1\lesssim n_{\rm H}\lesssim3000\cc$ even if this distinction is reduced in magnitude with weaker feedback. At higher densities ($n_{\rm H}>10^{4}\cc$) both simulations begin to build up a reservoir of gas waiting to be converted into stars. This build up could be the result of the simulation resolution we explore below.\\\\

Combing the result from \S\ref{fb:sfh} and \ref{fb:gas} (and as found by G17), it is clear that feedback does play an important role in shaping the structures within a galaxy. However, here we have also shown that while feedback is important so too is the location and conditions under which stars form. Indeed the location of star formation may be more important as this sets the location and conditions in which feedback is injected into the gas as well as temporal spacing between large feedback events (supernova). In short the difference found between MSF and TSF in \S\ref{res:LSF} is \emph{created} by the difference in star formation prescription and is then \emph{amplified} by the feedback (to what degree is set by the feedback model).

\subsection{Dependance on Resolution}
\label{res:resol}
%----- GMC Properties  -----%
\begin{figure*}
	\begin{center}
		\includegraphics[width=1.0\textwidth]{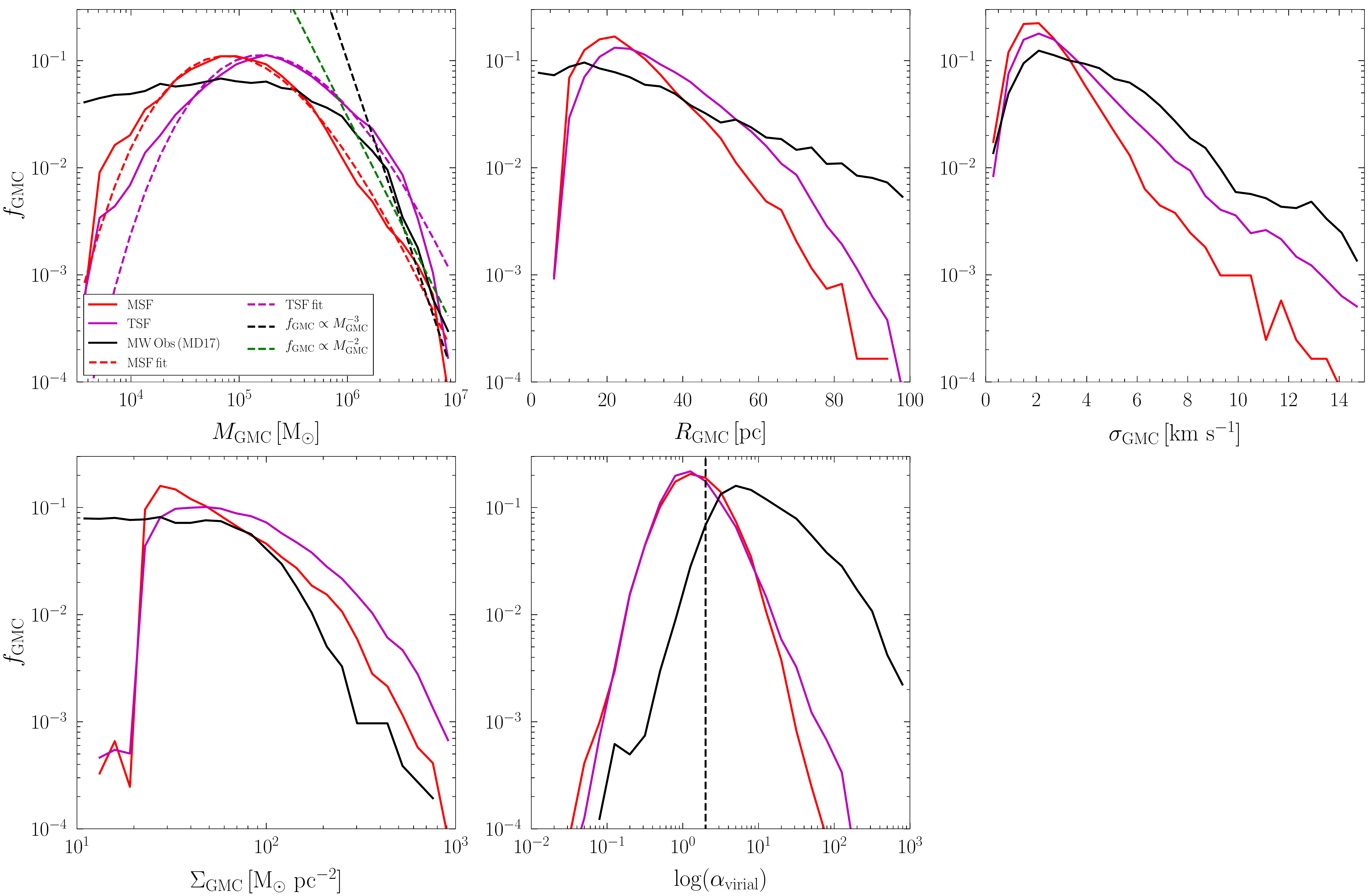}
		\caption{Normalised distributions of GMC properties ($\mgmc$, $\rgmc$, $\siggmc$, $\Siggmc$ and $\ap$). The red, magenta and black lines show data from MSF, TSF and data from MD17 clouds respectively.  The dashed magenta and red lines in the $\mgmc$ panel show log-normal fits to the TSF and MSF cloud distributions respectively, while the black and green dashed lines show power laws with indexes of $-3$ and $-2$ respectively, see \S\ref{res:2D} for details. 
		The dashed vertical line in the $\ap$ panel shows $\ap=2$, i.e the border between bound and unbound.
		}
	\label{fig:clouds}
	\end{center}
\end{figure*}
%----- ---------------------- -----%

To determine the extent that numerical resolution impacts our results we have rerun both MSF and TSF (using the fiducially feedback model) at several different resolutions: $\sim18.3,\,9.6$ and $2.3\pc$ and carried out the same analysis as above on these simulations. In this section we present the conclusion of this analysis, but provide the figures showing the \sfr, stellar TPCF and gas PDF at different resolutions in Appendix \ref{app:res}. 

The resolution of simulation largely sets the maximum density that gas in our galaxies is able to reach, with higher resolution allowing for higher densities. This in turn impacts when, where and the rate at which star particles are formed: i.e.  increasing resolution increases the maximum $n_{\rm H}$ and fraction of gas found at $n_{\rm H}\gtrsim10\cc$ which leads to a higher \Sfr. As TSF requires gas to be self-gravitating before a star particle can be created and gas density plays an integral factor in determining if a cell is self-gravity, TSF is less efficient at forming star particle in the $\Delta x\sim18.3$ and $\sim9.6\pc$ simulations. This results in ratio of total stellar mass formed in TSF compared with MSF dropping from $1.05$ to $0.86$ and $0.933$ in the $\Delta x\sim18.3$ and $\sim9.6\pc$ runs respectively. The ratio of formed stellar mass in TSF to MSF increases to $\sim1.25$ when $\Delta x\sim2.3\pc$ which suggest that our finical resolution is the closest the simulations will get to converging. 
The TPCFs for the different resolution simulations likewise shows that for a given resolution TSF has a more clustered star formation than MSF and maintains approximately the same power law index for all resolutions. 

The results in \S\ref{fb:sfh} show that our fiducial feedback model is able to regulate the star formation history of a simulated galaxy to $\sim6\Msol\yr^{-1}$, and the results of the lower resolution runs seem to support this: $M_{\star,{\rm tot}}$ for different resolutions are approximately parallel to our finical resolution for both MSF and TSF. This is not the case in the high resolution TSF simulation. In TSF a large number of stars are injecting feedback (particularly SNe) into high density regions. From conservation of momentum arguments it is possible to calculate the magnitude of the velocity injected into these gas cell by each SN event. Due to the high density ($n_{\rm H}>10^{4}\cc$) and small cell size, for the $\Delta x\sim2.3\pc$ simulation the injected velocity is $\lesssim1\kmsec$ and thus unable to have a meaningful impact on the gas. From the PDFs (see Fig.~\ref{fig:resgpdf}) we know that MSF has less gas at $n_{|rm H}>10^{4}\cc$ than TSF at all resolutions and is therefore less susceptible to this issue. If these simulations are to be pushed to higher resolution it is likely that the feedback model needs to be reexamined to ensure that sufficient momentum is inject into the ISM or that additionally feedback mechanisms, i.e. magnetic fields, are needed. 

In summary, simulation resolution sets the maximum gas density and final stellar mass of the simulations when run with a maximum resolution of $\Delta x\geq4.6\pc$. At higher resolution the turbulent star formation prescription forms a significant fraction of its star particles in regions where feedback injects insufficient momentum to be efficient. 

The resolutions explored by our tests suggest that convergence between MSF and TSF occurs when the simulations have both have resolutions around $\Delta x\sim4.6\pc$ (as discussed above), however convergence between the different resolution runs of the same simulation is not reached. In the case of MSF convergence is only likely to occur when the formation of single stars is modelled rather than star particles. TSF, on the other hand, is likely to converge when sub-sonic turbulence is resolved (i.e. $\Delta x <0.1\pc$). In the sub-sonic regime both $\sigma_{s}$ and $s_{\rm crit}\rightarrow0$ and Eq.~\ref{eq:sfrff} becomes 
\begin{equation}
	\effsf=\frac{\epsilon_{\rm PS}}{2\phi_{t}} = 0.1425,
\end{equation}
and no longer depends on the velocity of the cell. In this regime star formation set simply by the gas density and value of $\Msi$, making the turbulent star formation model almost identical to the molecular star formation model. 

%----- GMC Properties  -----%
\begin{figure*}
	\begin{center}
		\includegraphics[width=1.0\textwidth]{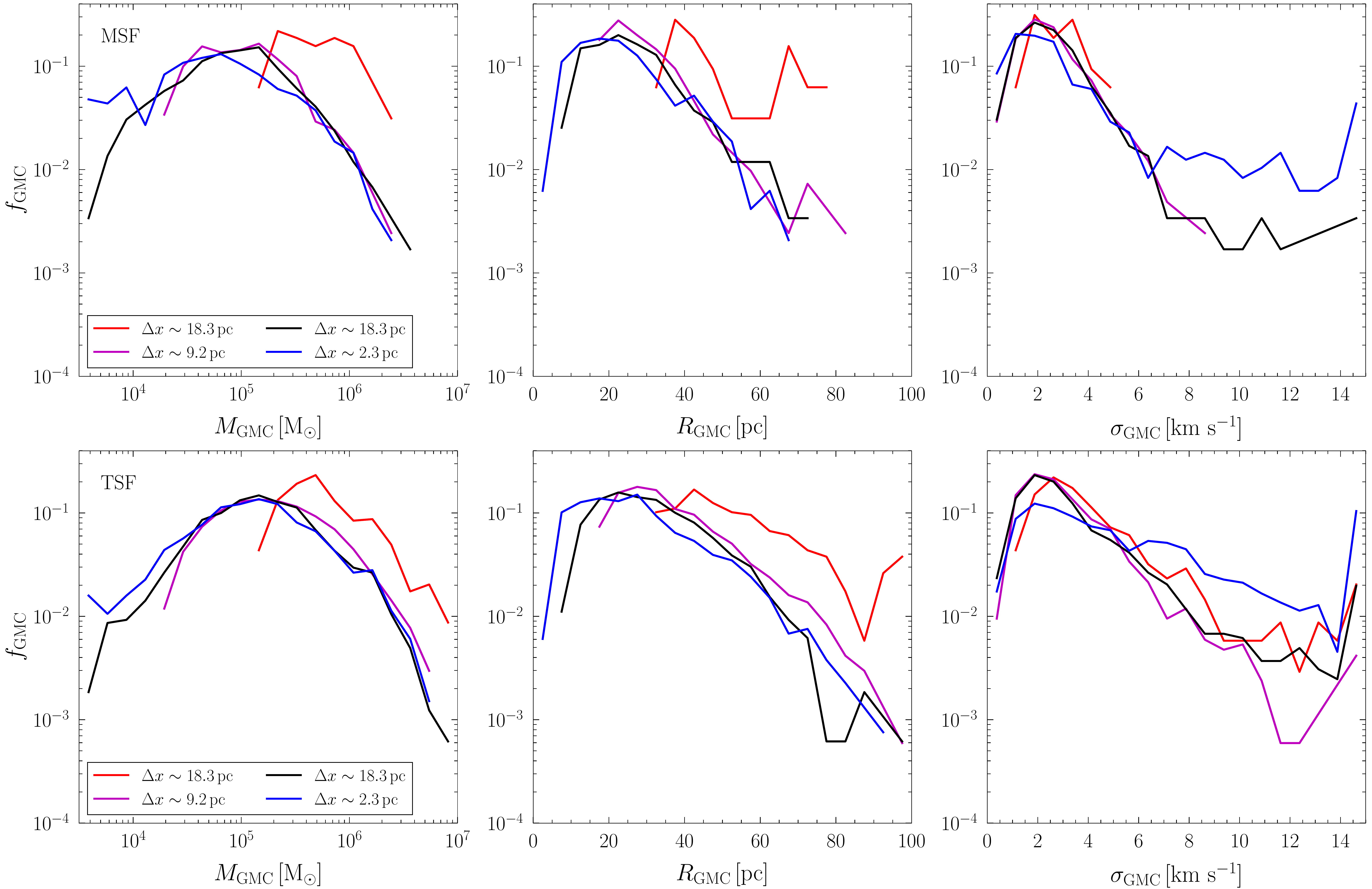}
		\caption{ Comparison of the normalised distributions of GMC properties ($\mgmc$, $\rgmc$, $\siggmc$) for simulations run at different resolutions. The top panels show the distributions for clouds identified in MSF simulations while the bottom row shows clouds from TSF simulations. Each panel shows the results from their respective simulations run with a maximum resolution of $\Delta x\sim2.3,\,\sim4.6,\,\sim9.2$ and $\sim18.3\pc$ (blue, black, magenta and red lines respectively). NB: Cloud finding parameters have not been changed, see \S\ref{meth:gmcs} for details. Only clouds identified at $t=450\Myr$ are included in this figure. 
		}
	\label{fig:cloudsres}
	\end{center}
\end{figure*}
%----- ---------------------- -----%

%----- 2D Larson Relations  -----%
\begin{figure*}
	\begin{center}
		\includegraphics[width=1.0\textwidth]{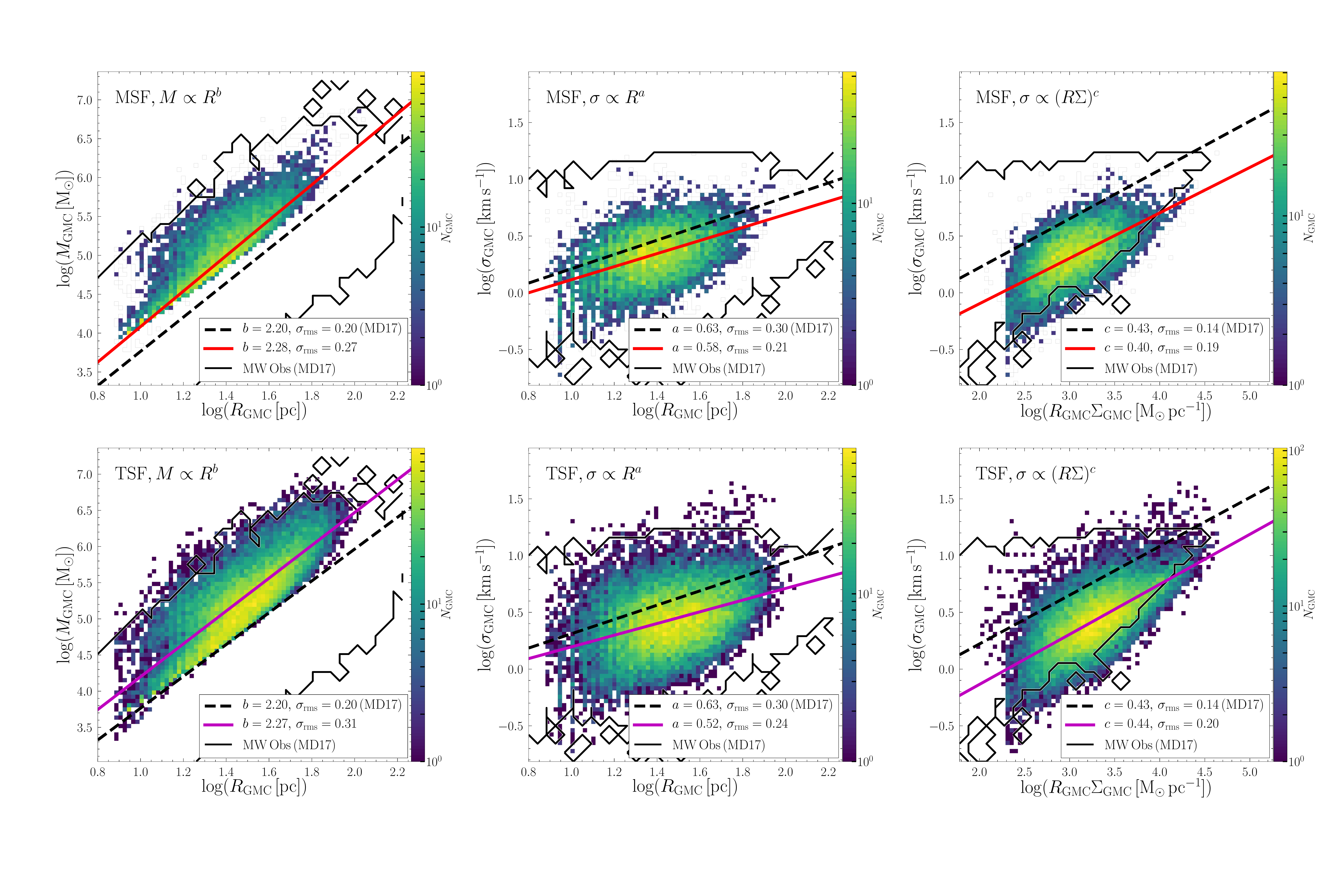}
		\caption{ 2D histograms comparing properties of GMCs. From left to right we show: $\mgmc$ as a function of $\rgmc$, $\siggmc$ as function of $\rgmc$ and $\siggmc$ as a function of $\rgmc\Siggmc$. The top row shows the data for GMCs found in MSF, while the bottom row shows GMCs in TSF. We fit a Larson-like relation for each panel (solid line) and show the relationship stated in MD17 (dashed line). The measured power law and the root-mean-square scatter ($\sigma_{\rm rms}$, in dex) of these two fits are given in each panel.  The black contours show the equivalent values for observed MW GMCs, as measured by MD17.
		}
	\label{fig:larson}
	\end{center}
\end{figure*}
%----- --------------------------- -----%

%--------------------------------------------------------------------------- Section: Results: 2D Clumpfinding -----------------------------------------------------------------
\subsection{GMCs}
\label{res:2D}

\subsubsection{GMC Properties}
\label{res:2Dgen}
Applying the clump finding method to MSF and TSF results in 12,149 and 23,777 GMCs being identified respectively over the $300\Myr$ period of analysis. In Fig.~\ref{fig:clouds} we show the normalised distribution of: molecular gas mass ($\mgmc$), radius ($\rgmc$), velocity dispersion ($\siggmc$), surface density ($\Siggmc$) and the virial parameter ($\ap$) for these GMCs (see \S\ref{meth:gmcs}). The figure also shows the properties of GMCs observed in the real Milky Way as reported by \cite{Miville-Deschenes:2017aa}, henceforth MD17. As noted in G18, due to the resolution of the simulation and our choice of clump finding parameters we fail to capture low mass ($\mgmc\lesssim10^{2.5}\Msol$) and compact ($\rgmc\lesssim7\pc$), however such clouds are found in the MD17 catalogue. 

Comparing MSF to TSF, we see a systematic shift to larger values for both $\mgmc$ and $\rgmc$. Given that the PDF of TSF showed an increased fraction of gas at medium to high densities (i.e. $10\lesssim n_{\rm H}\lesssim10^{4}\cc$), this increase is not unexpected. The larger fraction of gas at these densities leads to more (molecular\footnote{As in G18 and G19 we assume that all gas with $n_{\rm H}\geq100\cc$ is molecular.}) gas above the density threshold used for identification. Furthermore, this section of the PDF is a continuous range of densities, therefore not only will the most dense region have a larger density but so too will the surrounding regions which leads the clump finder to find larger clouds. A visual inspection of several clouds along a spiral arm in TSF found them to be more extended (in the direction parallel to the arm) than those in the equivalent arm of MSF. The more extended clouds will thus have a larger area which leads to a large $\rgmc$ and $\mgmc$. From $\Siggmc$ we see that clouds which are more massive also tend to be more extended. 

As the feedback model employed in these simulations drives the gas turbulence, or at the very least reduces the fraction in solenoidal motion (see Fig. 13 of G17), similarity in $\siggmc$ for the two simulations is also expected. While, both simulations produce clouds with $0\leq\siggmc\leq15\kmsec$ we find more clouds at higher $\siggmc$ in TSF, for example $\sim0.08\%$ of clouds have $\siggmc=14.1\kmsec$ in MSF but this rises to $6.3\%$ in TSF. This increase in $\siggmc$ is mostly likely a result of the increased clustering in star formation (see \S\ref{res:LSF}). As $\ap$ depends on $\siggmc^{2},\,\rgmc$ and $\mgmc$ (see Eq.~\ref{eq:virial}), the increase in the former two is effectively cancelled by the increase $\mgmc$. This results in the distribution of $\ap$ being nearly identical for the two simulations. 

We now compare the simulations to the observations and find reasonable match between TSF and MD17, however this match is far from perfect. As mentioned above, our simulations lacks low mass and small clouds, which creates a discrepancy between them and observations. If we remove clouds from the observational data set that are below the simulations' $\mgmc$ and $\rgmc$ limit, we do find a much better match with observations but there is still room for improvement. In particularly, the miss match in $\Siggmc$ in TSF and MD17 is almost entirely resolved when these clouds are removed from the comparison. 
For both $\rgmc$ and $\siggmc$ the simulations have less clouds at high values and in particularly MSF is missing a significant fractions of large radius and high velocity clouds. 

It is worth highlighting that the method used to identify GMC and extract their properties from our simulations is not a perfect match to the methods used in MD17. For example, MD17 made use of velocity measurements when extracting clouds and required specialised selection functions due to line of sight dependant linear resolution. The differences in cloud identification method could therefore provide an explanation for discrepancy between our GMC data and those of MD17. 

MD17 parameterise the high mass end of their $\mgmc$ distribution as power law with and index of $-3$  (see their \S4.1.3 and Fig.~7)\footnote{MD17 state the power law  in their work as ${\rm d}N/{\rm d}\ln{\mgmc}\propto \mgmc^{-2}$, which is equivalent to ${\rm d}N/{\rm d}\mgmc\propto\mgmc^{-3}$}. For our simulations we find that the high mass end is better fit with a power law index of $-2$.  The entire $\mgmc$ distribution, for both simulations, is best described by the log-normal distribution,
\begin{equation}
	f=\frac{a}{x\sigma\sqrt{2\pi}}\exp{\left(
	-\frac{(\ln{x} - \mu)^{2}}{2\sigma^{2}}    
	\right)},
\end{equation}
where $a$ is scaling constant, $x=\log_{10}(\mgmc)$, and both $\mu$ and $\sigma$ are fitting parameters. TSF (MSF) is best fit by $\mu=1.66$ ($1.59$) and $\sigma = 0.095$ ($0.1$). It is possible that MD17's catalogue of GMCs might also be described by a log-normal, however in this case a power law tail would be required for $\mgmc\lesssim10^{5}\Msol$. 
It is worth noting that \cite{Colombo:2019aa}, like MD17, found ${\rm d}N/{\rm d}\mgmc\propto\mgmc^{-1.7}$ for GMCs with mass between $\sim10^{3}$ and $10^{6.5}\Msol$ in the MW.  Also, as with MD17, the detection and selection methods used in \cite{Colombo:2019aa} are different to those used in this work, which could be the cause of the discrepancy between our log-normal fit and the power law fits  found in observations. 

It is also possible to compare the GMC property distributions to extragalactic observations. For example, \cite{Grasha:2018aa} shows the distributions of $\mgmc,\,\rgmc$ and $\siggmc$ for NGC 7793. The general shapes of these distributions appear to be similar to those presented in Fig.~\ref{fig:clouds}, i.e. $\mgmc$ appears to follows a log-normal distribution while both $\rgmc$ and $\siggmc$ peak at small values ($\sim10\pc$ and $\sim2.3\kmsec$ respectively) with the number of clouds at larger values decreasing.  Its worth noting that all three of NGC 7793's cloud properties are shifted to smaller values than our galaxy. van Donkelaar \& Grisdale (in prep.) has found that in smaller LMC-like galaxies GMC properties tend to be similarly shifted to lower values. Thus the it is not unexpected that NGC 7793, which has approximately a tenth of the gas mass of our simulated galaxies \citep[][]{Saikia:2020aa} has distributions centred at lower values. \\

\subsubsection{Resolution and GMC properties}
\label{res:2Dres} 
As discussed in \S\ref{res:resol}, we have run our simulations with four different maximum spatial resolutions, which allows for a determination of how resolution impacts the properties of GMCs identified within each simulation. Fig.~\ref{fig:cloudsres} shows the clouds identified in a single simulation snapshot ($t=450\Myr$) for each resolution and star formation method. We find that the property distributions of clouds in the TSF simulations are less susceptible to the resolution of the simulation than clouds found in MSF.

For both TSF and MSF, when moving from high ($\sim2.3\pc$) to low ($\sim18.2\pc$) resolution the minimum radius of cloud increases as does its minimum mass.  This is expected behaviour: our cloud identification parameters require at least nine cells per cloud so as $\Delta x$ increases so too does the minimum size of cloud. Likewise, by preventing cells from being refined mass builds up on the smallest allowed scale. Increasing the size of clouds also results in less clouds being identified as the reduced resolution causes neighbouring clouds to be merged in to single objects. This is particularly true when using $\Delta x\sim18.3\pc$, where only 32 clouds are identified in MSF, compared with 412 for $\Delta x\sim9.2\pc$ and hence why $\Delta x\sim18.3\pc$ data points in Fig.~\ref{fig:cloudsres} are normally found at higher $f_{\rm GMC}$ compared to the other resolution runs. 

What is not necessarily expected is the very limited range of $\siggmc$ found in MSF with $\Delta x\sim18.3\pc$ and also slightly limited range of $\mgmc$ and $\rgmc$ as well. This appears to be the result of where feedback is introduced into the simulation. As discussed above, star formation (and therefore feedback) occurs in any cell with $\geq3333\Msol$ of gas in MSF and it is therefore possible for stars to form outside of GMCs at low resolutions. These stars are able to remove loosely bound gas from outer edge of GMCs leaving only the gravitationally bound gas and thus leading to a smaller range of $\siggmc$. In TSF, however, stars only form inside GMCs so all feedback energy is injected into the interior of clouds which increases their $\siggmc$. 

\subsubsection{Larson Relations}
\label{res:larson}
Fig.~\ref{fig:larson} shows the Larson relations ($\mgmc\propto\rgmc^{b},\,\siggmc\propto\rgmc^{a}$ and $\siggmc\propto(\rgmc\Siggmc)^{c}$) for MSF and TSF. As in G18, we find excellent agreement between the power law indexes (i.e. $a,\, b$ and $c$) measured from our simulations and those measured from MD17 (see figure for values). Given that the simulation without feedback run in G18 failed to match the observed relations and that both MSF and TSF do, we argue that these relations are not dependant on the how and where stars form within the GMC, but do depend on the feedback from stars to (at least partially) drive GMC evolution so that clouds lie on these relations. Finally, we note that the distribution of clouds in $\mgmc$-$\rgmc$, $\siggmc$-$\rgmc$ and $\siggmc$-$\rgmc\Siggmc$ parameter spaces is more extended for TSF than MSF, which is a result of the increase in the number of high mass and large radii clouds. These more extended the distributions of TSF are a better match for the distribution from the MD17 catalogue.

As in G18, Fig.~\ref{fig:clouds} shows the distribution for all clouds identified over the $300\Myr$ period of analysis. However it is important to consider if and how this distribution evolves with time. The mass distribution, for example, is nearly identical at all time-steps but with small variations in the number of clouds with a given mass. This behaviour is seen in all properties. The maximum standard deviation from the time-averaged mean is $\sim0.1$ for TSF and $\sim0.2$ for MSF, i.e. their is marginally less variation in cloud properties distributions in the former. We also find little variation in the measured values of $a,\, b$ and $c$, suggesting that relationship between properties is also largely time-invariant.

%<<<<<----- Figure: eff ----->>>>>%
\begin{figure}
	\begin{center}
		\includegraphics[width=0.5\textwidth]{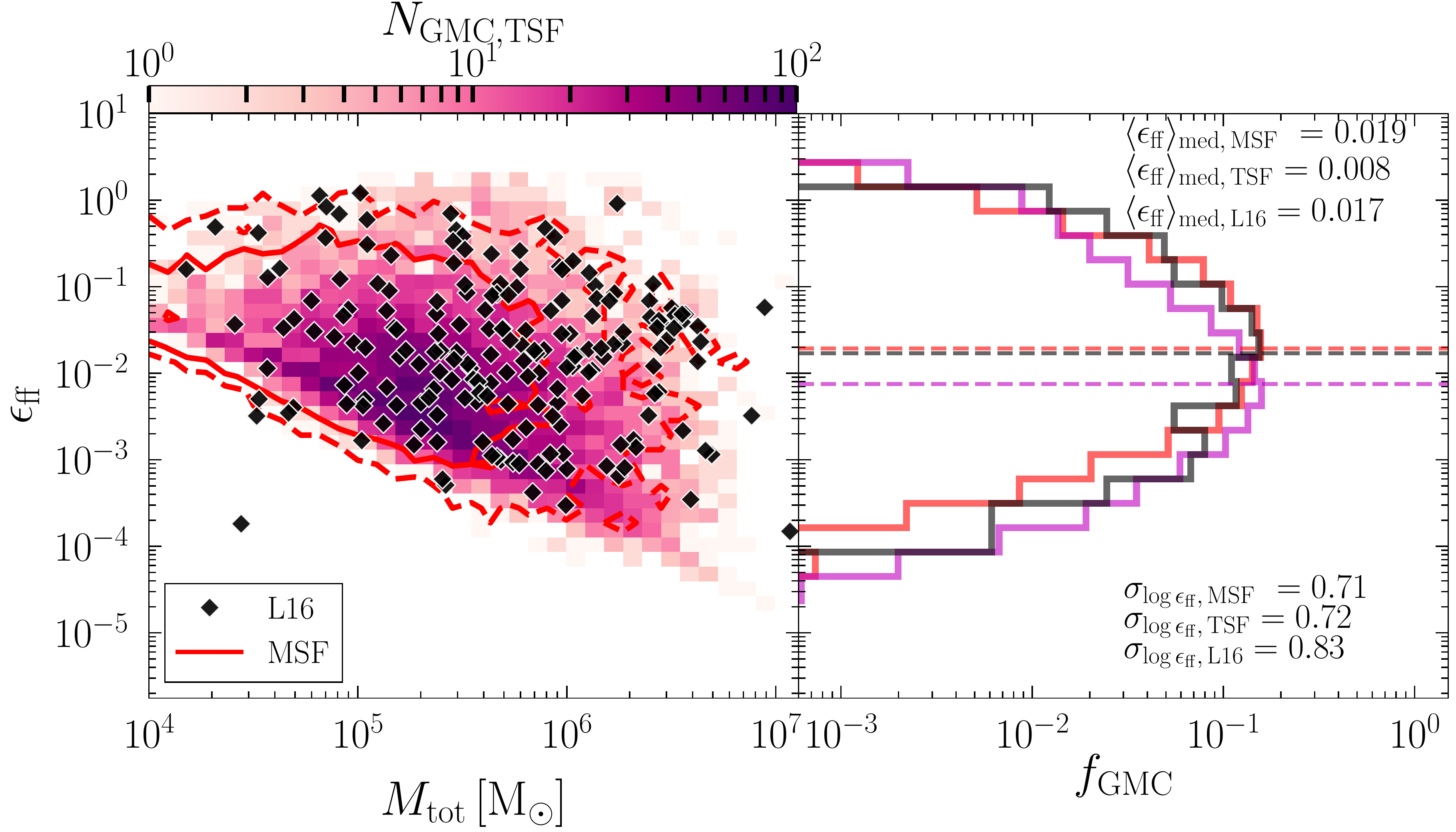}
		\caption[]{Left: $\eff$ as function of total cloud mass ($\mtot$). GMCs from TSF are shown by the magenta 2D-histogram, while those from MSF are shown by the red contours ($\geq1,\,\geq10$ GMCs). Right: Histograms showing the fraction of GMCs ($f_{\rm GMC}$) with a given $\eff$. GMCs from  MSF are in red, while those from TSF are shown in magenta. 
		 The dashed lines show the median efficiency ($\sfepmed$) for each data set. The values of the $\sfepmed$ and the standard deviation in $\log{\epsilon}$ ($\sigma_{\log\epsilon}$) are given in the panel. 
		 For comparison with observations, data from Table 3 of L16 is included in both panels (black points and black histogram respectively) and $\langle\eff\rangle_{\rm med,\,Lee16}$ is shown (dashed black line).
		} 
		\label{fig:eff}
	\end{center}
\end{figure}
%<<<<<----- --------------------------- ----->>>>>%

%--------------------------------------------------------------------------- Section: Results: eff  -----------------------------------------------------------------------------------
\subsection{Distribution of Star Formation Efficiencies}
\label{res:eff}
We now turn our attention to star formation within the GMCs. As discussed in \S\ref{sect:intro}, measurements of how efficiently GMCs convert gas into stars produces a range of values, which depends not only on the cloud being observed but also on the technique being used to observe them. We adopt Eq.~\ref{eq:sfeff}, as our estimator for the efficiency of gas to star conversion, which is the same estimator used in G19 and \cite{Lee:2016aa} (henceforth L16) to which we compare our results. The left panel of Fig.~\ref{fig:eff} shows how $\eff$ of clouds varies as function of total mass ($M_{\rm tot}=\mgmc+\msy$) and find almost identical results to G19, i.e. a scatter in $\eff$ at all values of $M_{\rm tot}$ as well as a reasonable match in the ranges of both $M_{\rm tot}$ and $\eff$ to L16. The $\eff-M_{\rm tot}$ panel seems to indicate that smaller GMCs are more efficient at forming stars, however as G19 discusses, this could be the result of a sampling bias or limit in both observations and the simulations. As exploring such a bias is beyond the goals and scope of this work we do not comment further on this potential relation. 

The right panel of Fig.~\ref{fig:eff} shows the fraction of clouds with a given $\eff$, as well as the median value ($\effmed$) and the standard deviation from the median ($\sigeff$\footnote{$\sigeff$ is robustly estimated via the median absolute deviation (MAD): $\sigeff= 1/0.6745$ MAD \citep{Muller:2000aa,Romeo:2016aa}.}). Both MSF and TSF are in good agreement with L16. Additionally, $\sigeff$ for both MSF and TSF are almost identical ($0.71$ and $0.72$ respectively) and only slightly smaller than the value we measure from L16 ($0.83$). In the case of MSF, the difference in $\sigeff$ is a result of missing the very inefficient clouds ($\eff\lesssim2\times10^{-4}$), while for TSF it is the reduced number of clouds with $0.02\lesssim\eff\lesssim2$. This results in a slight reduction in $\effmed$ to $0.008$ for TSF, which is still within a factor of two of the value we measure form L16.

Due to the fixed $\effsf$ (see \S\ref{sims:kmt}) used in MSF a large fraction of each GMC will be star forming. This in essence sets a minimum value for $\eff$\footnote{Feedback complicates this due to the removal of gas from small scales and thus raising the minimum measured value of $\eff$.} in the MSF simulation. TSF has an $\effsf$ that varies from cell to cell and as a result cells that would form stars in MSF (because of their mass) don't necessarily in TSF. This allows for regions within a GMC that do not form stars and thus allowing for a smaller minimum measured value of $\eff$ (when measured over an entire GMC).

%-------------------------------------------------------------------------------------------------------------------------------------------------------------------------------------------
%--------------------------------------------------------------------------- Section: Discusion --------------------------------------------------------------------------------------

\section{Discussion}
\label{sect:Disc}

%--------------------------------------------------------------------------- Section: Discussion: Galactic Scales --------------------------------------------------------------
\subsection{Comparison with Previous Studies}
\label{disc:gs}

\subsubsection{Galactic Scales}
There are several star formation methods employed in the simulation community to determine when and where star formation should occur. Most of these methods have shown to have success in matching observations in various ways. For example, \cite{Semenov:2018aa} simulated an isolated $L_{\star}$-like galaxy, employing a self gravitating criteria to determine when and where stars should form. Similar to our MSF simulation, during run time all cells in the simulation have the same density criteria and value of $\effsf$. Their fiducial model, using $\alpha_{\rm vir}<10$ as their condition for star formation and  $\effsf=1\%$, was able to match ``all considered observations reasonably well''. Furthermore, they showed that with sufficiently large enough $\effsf$ (or strong enough feedback) that the galaxy enters a self-regulation regime, similar to what we found for MSF with $\effsf=100\%$.

As described in \S\ref{sect:intro}, H13 found that the morphology (i.e. spatial and density distribution of gas) of a simulated galaxy ``depend strongly on the star formation criteria'' used. With regards to their two simulations using similar star formation criteria to those used in this work, H13's isolated MW-like galaxies are similar in some aspects (e.g. the most prominent spiral arms appear to have nearly identical position, pitch angle and length) but different in others (number of arms, structure in the galactic centre, etc.).  At first glance this seem to be in slight contrast with the simulations presented in this work, i.e. both MSF and TSF have remarkably similar morphologies However we argue that this discrepancy can be explained by the differences between the star formation methods used in this work and H13. 

The prescription used in the MSF simulation is very similar to  H13's ``Molecular Gas'' method (criteria 3 in their paper), with one subtle but important difference: the choice of $\effsf$.\footnote{There are of course other differences between the simulations such as: simulations codes ({\sc gadget} vs. \ramses), feedback implementations, $M_{\rm \star,i}$, etc. However, given how similar their MW simulation with Self-Gravity criteria is to our MSF and TSF models we argue that these differences are significantly less important than how the star formation criteria is implemented. } 
This work uses $\effsf=10\%$, which results in an initial burst of star formation (at $t<50\Myr$), followed by feedback reducing and regulating the \sfr, resulting in a \emph{measured} $\eff\sim1\%$ \citep[Fig.~\ref{fig:eff}, see also][ and G19]{Agertz:2013aa,Agertz:2015aa}. By contrast H13 uses $\effsf=1.5\%$ (labelled at $\epsilon$ in their work) while this does give rise to a similar (though smaller) initial burst of star formation, their SFH is markedly different to MSF: their \sfr steadily rises from $\sim0.7$ to $\sim5\Msol\yr^{-1}$ over $\sim650\Myr$. We suspect that this is likely the result of their simulation not producing stars fast enough for their their feedback model to efficiently regulate the \Sfr. 

Their ``self-gravity'' method (criteria 1 in their paper) also has notable differences to our turbulent star formation model. Most notably they employ a fixed $\effsf=100\%$ and assume a single free-fall time for any gas that is self gravitating. Our turbulent star formation prescription does not explicitly set $\effsf$, instead the model allows each computational cell to have a unique $\effsf$ of any value, which is calculated at each time step, see  \S\ref{sims:turb}. It is therefore possible for a gas cell to have an $\effsf>100\%$ (see \S\ref{res:LSF} for the range of $\effsf$ measured during run time). Furthermore, the model we employ is specifically designed to allow each cell to have a unique free-fall time \citep[see discussion in \S2.4.5 of ][]{Federrath:2012aa}. These differences combine to result in the \sfr of TSF being larger at almost all times compared with the self-gravity model presented in H13. It maybe interesting to rerun our simulations using a self-gravity criteria identical to that presented in H13, however we leave this for future work. 

Despite the differences in morphologies, H13 found that different star formation prescriptions resulted in ``identical integrated star formation rates''. As discussed above (\S\ref{res:gp}) and shown in Fig.~\ref{fig:sSD} we also find that the choice of star formation criteria has little impact on the total stellar mass over a period of $450\Myr$. This result combined with the \sfr of our simulated galaxies (see Fig.~\ref{fig:sfh}) appears to confirm H13's conclusion: feedback is the most important factor in determining the \sfr of a galaxy not star formation criteria. 

In addition to the isolated MW-like case discussed above H13 also compared the results of different star formation criteria in galaxies with ISM conditions similar to those found in a merger. In these conditions the morphological differences arising from the `self-gravity' and `molecular' star formation prescriptions are significantly more pronounced. The Horizon collaboration carried out cosmological simulations using both a simple density based star formation criteria as well as a (nearly) identical turbulence criteria as the one used in this work. They found that the different star formation models produced significantly different results for galaxies under going mergers (private communication). This is largely a result of a merger producing regions of high density but with a divergent velocity field. While the density based method formed stars in these regions,  the turbulent method did not. This result suggests a possible observational test of which method best matches how the Universe actually form stars.

The \newhorizon cosmological simulation \citep[][]{Dubois:2020aa} uses the same turbulent star formation prescription implemented in this work. \newhorizon has been been able to reach fair agreement with the observed Kennicutt-Schmidt relation suggesting that such a star formation method produces ``realistic'' galaxies when employed in a cosmological context even when resolution is limited to $\sim30\pc$.

\subsubsection{GMC Scales}
One of the most popular methods to studying GMCs is the use of isolated cloud simulations. These allow for significantly higher resolution than simulations of entire galaxies and allow for the study of internal cloud structures \citep[e.g.][]{Audit:2010aa,Padoan:2016aa,Grudic:2018aa}. These simulations can be invaluable tools for determining the role of star formation and feedback within a cloud, for example, \cite{Grudic:2019ab} showed that in an isolated GMC varying the ultraviolet component of feedback lead to variations in the measured star formation efficiency of GMCs.  

The SImulating the LifeCycle of molecular Clouds (SILCC) project \citep[see][ for an over view of the project]{Walch:2015aa} has made use of magnetohydrodynamical simulations of a $0.5^{2}\times10\kpc^{3}$ volume to investigate, for example, how magnetic fields effect the chemistry within and formation of molecular clouds \citep{Girichidis:2018aa}. The SILCC project provides key insight into the processes that drive GMC and ISM evolution, particularly with respect to the inclusion of magnetic fields and ionising radiation (both of which are not included our simulations). However they neglect to include star formation and instead opt to include SNe by placing them randomly, in high density regions of the ISM or a mixture of the two. Furthermore, in general isolated cloud/ISM box simulations, such as those in the SILCC,  generally lack galactic context, i.e. the impact of neighbouring clouds, spiral arms, galactic rotation etc., making an assessment of how these process impact clouds difficult (some of these processes are often approximated). These limitations do not negate the usefulness of such simulations, indeed all simulations have limitations including the ones presented in this work, but an awareness of the missing physical processes is needed to probably compare results to both other simulations and observations. One method which may provide the resolution of the isolate cloud simulations with knowledge of the galactic environment is re-simulations. Here clouds are extracted from a galactic scale simulations and re-simulate at higher resolution \citep[e.g.][]{Rey-Raposo:2017aa}. In future work we aim to explore the potential of using this method.

%-------------------------------------------------------------------------------------------------------------------------------------------------------------------------------------------

\subsection{GMC Defintion}
\label{disc:settings}
The definition of GMCs is somewhat nebulous, with different studies (both observational and simulated) using different methods and criteria to identify them \citep[e.g. compare the methods of][]{Solomon:1987aa,Heyer:2009aa,Roman-Duval:2010aa,Dobbs:2014aa,Rice:2016aa,Miville-Deschenes:2017aa}. In this work we adopted the same definitions as those used in G18, which were tuned to make the distribution of $\mgmc$ of the MSF simulation similar to the distribution reported in \cite{Heyer:2009aa}. From a visual inspection of the clouds found in TSF and MSF we find in both that clouds tend to occupy the entire width of their host spiral arm but those in the former also extend further along the arm than those in the latter. Coupled with the fact that TSF also has an increase in the fraction of gas found at $10\lesssim n_{\rm H}\lesssim10^{4}\cc$, which is found primarily along the length of spiral arms, it appears that the criteria used to identify the boundary between clouds may be insufficient to separate different GMCs. 
Given the ambiguity in the definition of the GMCs, it would have be reasonable to re-tune our criteria on TSF. However such turning would limit our ability to compare to our previous work and perhaps more importantly this highlights the need for a specific definition of a GMC and how to define its boundaries. 

Additionally, as briefly discussed in \S\ref{res:2Dgen}, exacting GMCs from simulations and observations is often carried out in different ways. These different methods could lead one technique identifying an object as a cloud but another not. It is therefore important that when carrying out in-depth comparisons between clouds found in two data sets that the differences in method are well understood or accounted for at the least. Preferably it is would be advisable, in the case of simulations to match the observational methods as closely as possible.

Our definition of a GMC is further complicated by the simple assumption that all gas with $n_{\rm}\geq\rho_{\rm mol}$ is molecular (see \S\ref{meth:gmcs}). In practise molecular gas can be found in regions with density below $\rho_{\rm mol}$ \citep{Smith:2014aa} and depending on metallicity not all gas above $\rho_{\rm mol}$ might be molecular \citep{Gnedin:2009aa,Krumholz:2009aa}. To counter this limitation, in future work we will explore replacing the $\rho_{\rm mol}$ criteria by modelling the creation, advection and dissociation of molecular gas in the simulations. This will allow for cloud identification based on the molecular gas produced self-consistently within the simulation. 

Further complications with the definitions of GMCs arise from the resolution of the data set. As we showed in \S\ref{res:2Dres}, the resolution of the data used when carrying out cloud identification does have an impact on the number and properties of clouds. This is an issue for both observations and simulations which must be taken into consideration before comparisons between different data sets are carried out. 

%-------------------------------------------------------------------------------------------------------------------------------------------------------------------------------------------
%--------------------------------------------------------------------------- Section: Conclusion ------------------------------------------------------------------------------------
\section{Conclusion}
\label{sect:conc}
In this work we explored the impact of when and where stars form in a Milky Way-like isolated galaxy. The primary goal of this work is to determine if the choice of star formation law used in a simulation has a significant impact on the properties of a galaxy as a whole and in particular on  giant molecular clouds. To this end we employ two simulations each with a different star formation prescriptions. The first, MSF, allows all regions of the galaxy with molecular gas to form stars at fixed efficiency per free-fall time ($\effsf$). The second, TSF, only allows stars to form in regions that are self-gravitating with a varying $\effsf$ set by the turbulent motion of the gas. Our key results are as follows:
\begin{enumerate}
	\item Both star formation prescriptions produce galaxies with remarkably similar morphologies, both in terms of the gas and stellar structures. In particular, with the inclusion of feedback from stars, both methods produce almost identical spiral arms with only small variations in their position and angle. 
	
	\item The molecular star formation method is more efficient at converting gas into stars at densities of $10\lesssim n_{\rm H}\lesssim10^{4}\cc$. However, the turbulent method is more efficient at extreme densities ($n_{\rm H}>10^{4}\cc$) where gravity dominates over gas motions. As a result, TSF generally has a higher surface density along its spiral arms.
	
	\item While the overall morphologies of the galaxies are almost identical, the location of star formation is not! Both simulations primarily form stars along the spiral arms: MSF form stars along the entire length of each arm, while TSF forms stars in small high (gas) density pockets within the arms. Despite the different in the location of star formation, the general star formation histories of the two galaxies are remarkably similar, with the exception that the latter is generally more bursty. 
	
	\item By rerunning the last $130\Myr$ of both simulations but with different feedback we have been able to confirm that the difference between the TSF simulation and MSF simulation is a result of changing the star formation method and not the feedback. Feedback acts to either amplify or suppress the difference created by the different star formation methods. 
	
	\item The star formation prescription impacts the number GMCs found with a given gas mass ($\mgmc$), radius ($\rgmc$) and velocity dispersion ($\siggmc$), with clouds in the TSF simulation generally being shifted to larger values. The change in GMC properties is a direct consequence of the increase in the amount of gas found with $n_{\rm H}\geq10\cc$ combined with our choice of clump finding parameters. Furthermore, the distribution of GMC properties in TSF are a better match to those from observations and less affected by the resolution of the simulation.
		
	\item Despite the differences in the cloud properties, the relations between different properties (i.e. $\mgmc$-$\rgmc$, $\siggmc$-$\rgmc$ and $\siggmc$-$\Siggmc\rgmc$) remain unchanged. Both simulations produce relations consistent with the canonical Larson measurements. From this we conclude that these relations are not set by the properties of the cloud but are a consequence of the processes driving cloud evolution, e.g. feedback, cloud-cloud interactions and galactic environment. The distribution of clouds in the $\mgmc$-$\rgmc$, $\siggmc$-$\rgmc$ and $\siggmc$-$\Siggmc\rgmc$ parameter space is more extended in TSF than MSF, and better match to those found in observations. 
	
	\item Finally, by measuring the star formation efficiency per free-fall time ($\eff$) of GMCs we find that both simulations produce a median $\eff\sim1\%$ and a dispersion of $\sigeff\sim0.7{\rm\,dex}$ matching extremely well with observations (e.g. $1\%$ and $\sim0.8 {\rm\,dex}$ respectively).When combined with results from previous work, we conclude that this scatter and its size is not set by feedback or when/where stars are formed but is a natural consequence of diversity in properties and the evolution of clouds. 
\end{enumerate}

In summary, both star formation methods produce similar gas and stellar structures on large ($\gtrsim500\kpc$) scales while on smaller scales differences, such as an increase in star particle clustering and the heights density reached do vary. Despite the small scale difference both simulations produce GMCs that are a reasonable match to observations, with the TSF providing a better match to the high mass and large radius end of the cloud property distributions. Finally, when the resolution of the simulations are $\geq4.6\pc$ the turbulent star formation model shows less variation in both GMC properties and star formation histories. We therefore argue that the turbulent star formation prescription is the preferred method.

%-------------------------------------------------------------------------------------------------------------------------------------------------------------------------------------------
%--------------------------------------------------------------------------- Section: Acknowledgements -------------------------------------------------------------------------
\section*{Acknowledgments}
We thank the Erik Rosolowsky, our referee, for their valuable and insightful comments. 
We thank Julien Devriendt and Oscar Agertz for fruitful discussions.
KG acknowledge support from STFC through grant ST/S000488/1
This work was performed using the DiRAC Data Intensive service at Leicester, operated by the University of Leicester IT Services, which forms part of the STFC DiRAC HPC Facility (\url{www.dirac.ac.uk}). The equipment was funded by BEIS capital funding via STFC capital grants ST/K000373/1 and ST/R002363/1 and STFC DiRAC Operations grant ST/R001014/1. DiRAC is part of the National e-Infrastructure.

\section*{Data Availability}
The data underlying this article will be shared on reasonable request to the corresponding author.

%-------------------------------------------------------------------------------------------------------------------------------------------------------------------------------------------
%--------------------------------------------------------------------------- Section: Bibliography ----------------------------------------------------------------------------------
\bibliographystyle{mn3e}
\bibliography{ref}

%-------------------------------------------------------------------------------------------------------------------------------------------------------------------------------------------
%------------------------------------------------------------------------------------ Appendix -----------------------------------------------------------------------------------------
\appendix

%--------------------------------------------------------------------------- Appendix: Resolution Analysis -----------------------------------------------------------------------
\section{Resolution Analysis}
\label{app:res}
\begin{figure}
	\begin{center}
		\includegraphics[width=0.5\textwidth]{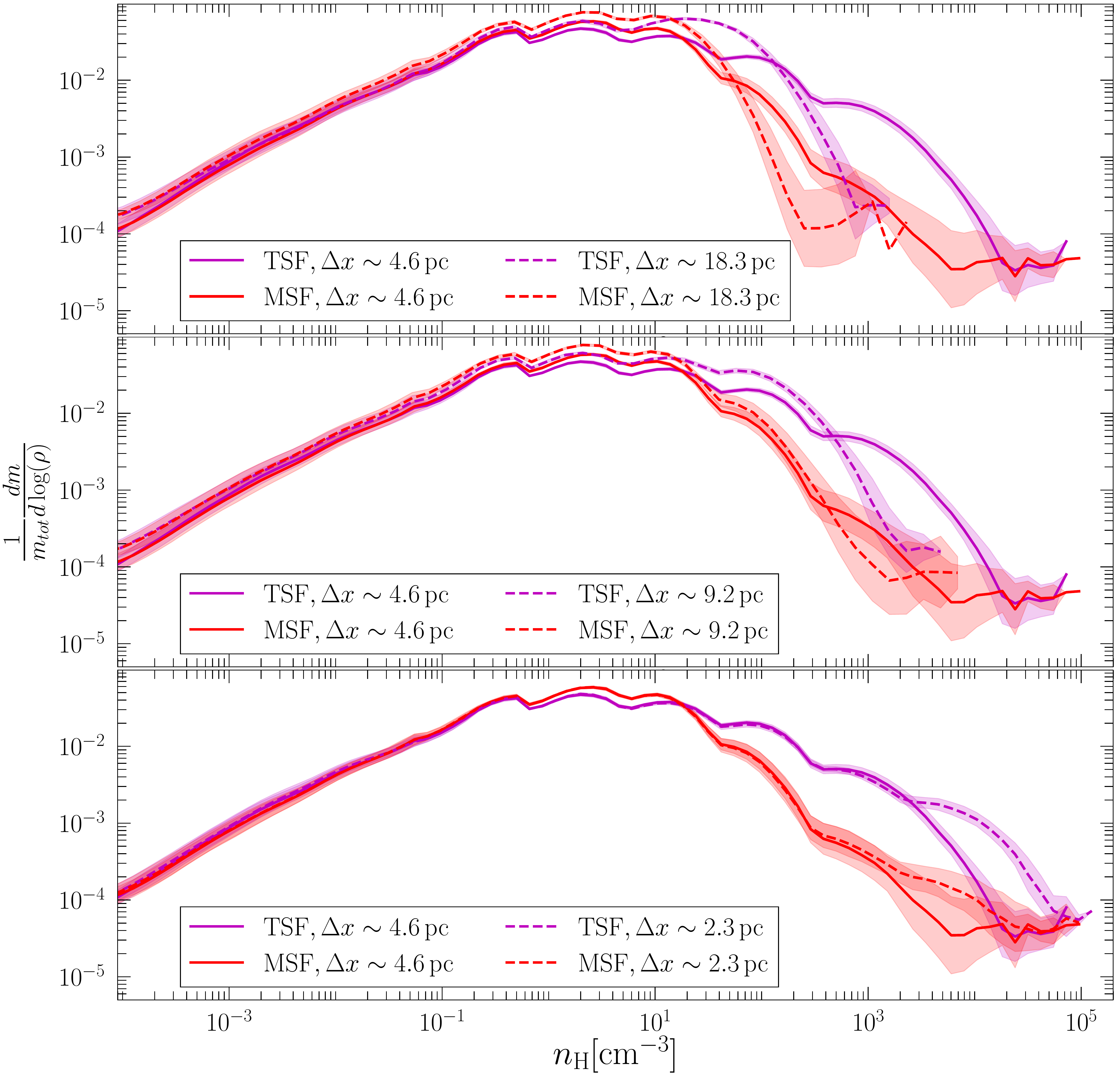}
		\caption{Comparison of the time averaged gas density PDF for our simulations at different resolutions. Each panel compares the fiducial resolution MSF and TSF simulations, i.e. $\Delta x\sim4.6\pc$ (solid red and magenta lines) which is compared to their equivalents at the resolution stated in the caption (dashed lines). Lines represent the average PDF for $150\lesssim t \lesssim 450\Myr$ and the shaded regions represent one standard deviation from the mean.
		}
		\label{fig:resgpdf}
	\end{center}
\end{figure}

\begin{figure}
	\begin{center}
		\includegraphics[width=0.5\textwidth]{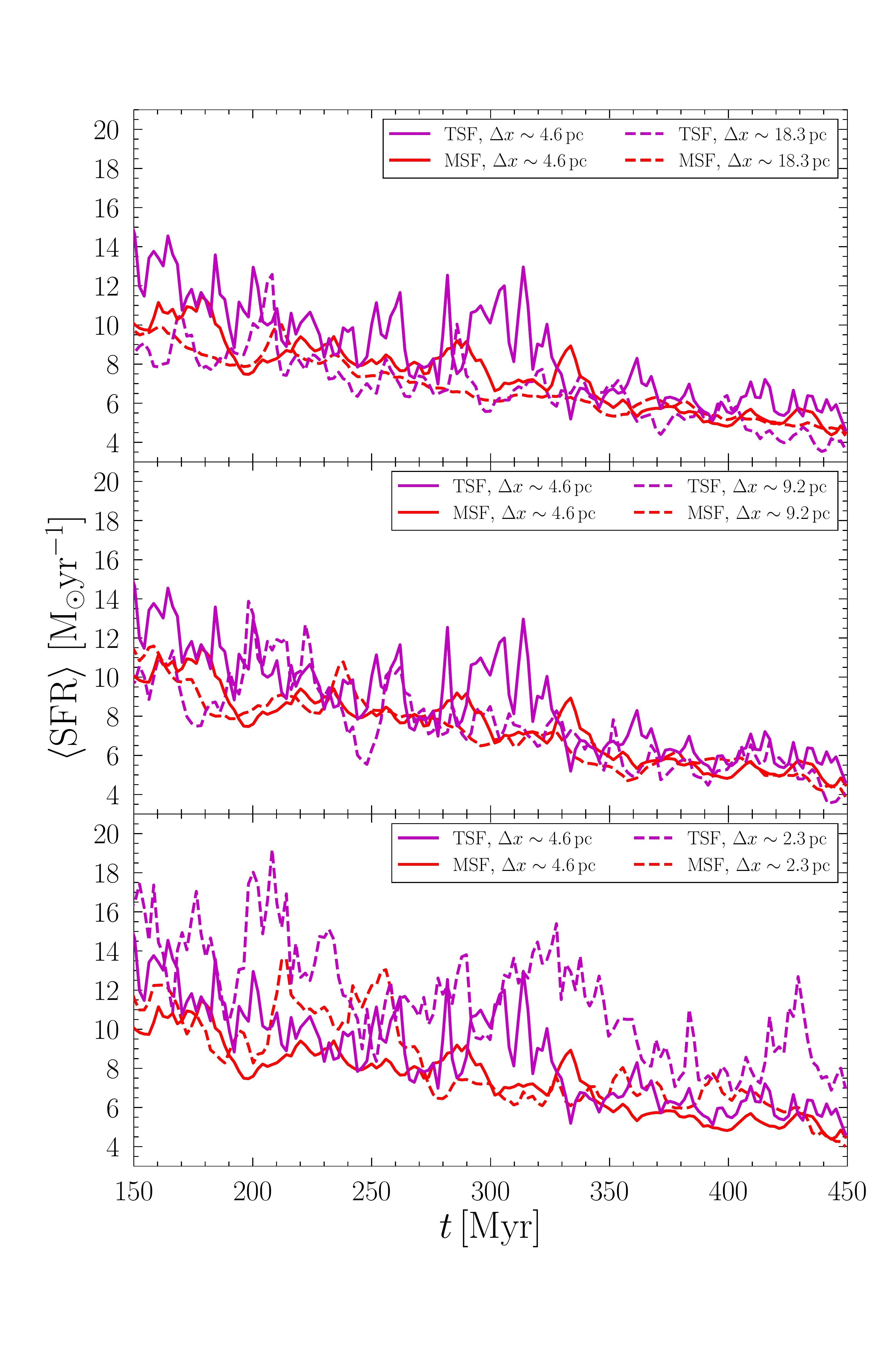}
		\caption{Comparison of the star formation rate (\Sfr) for Simulations run at different resolution. Each panel compares the fiducial resolution of MSF and TSF simulations, i.e. $\Delta x\sim4.6\pc$ (solid red and magenta lines) to their equivalents at the resolution stated in the legend (dashed lines).  
		}
		\label{fig:rescm}
	\end{center}
\end{figure}

\begin{figure}
	\begin{center}
		\includegraphics[width=0.5\textwidth]{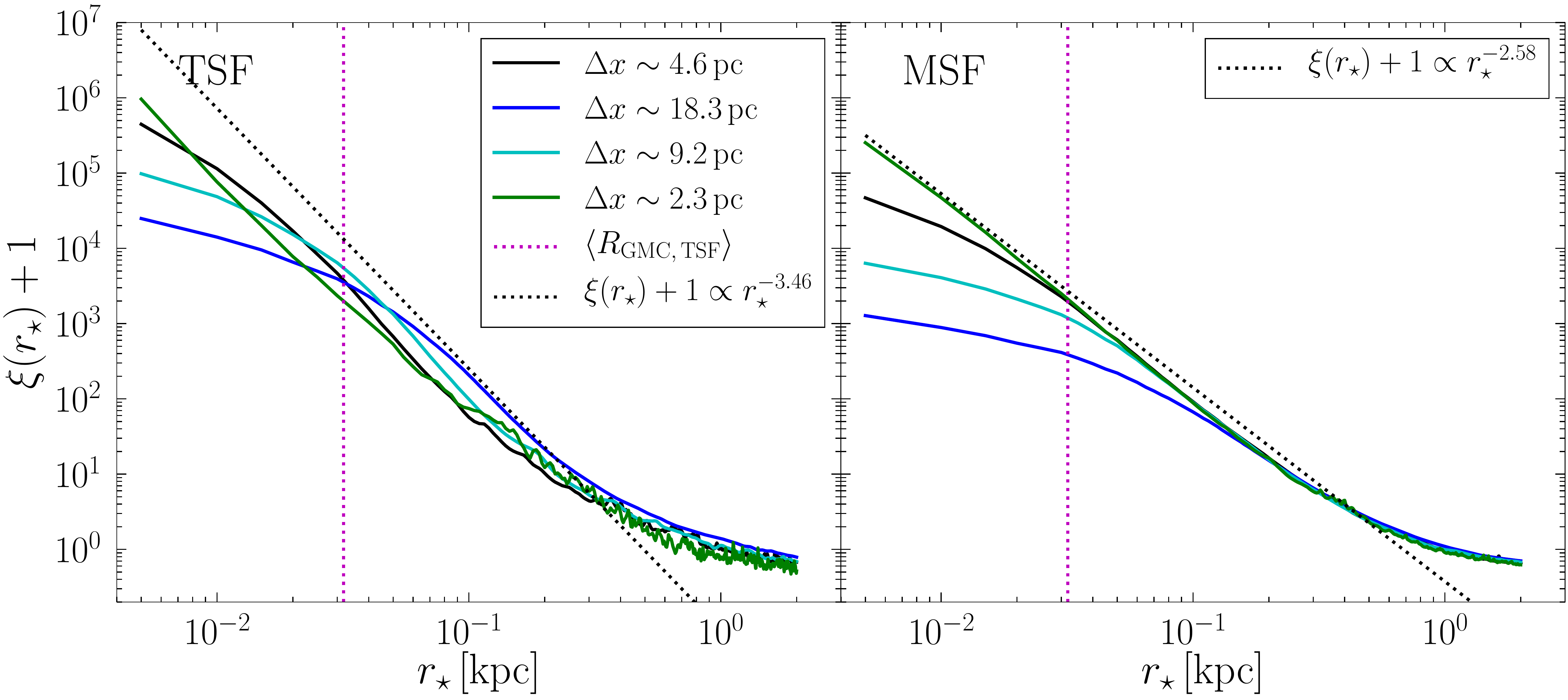}
		\caption{Comparison of the time averaged stellar TPCF for our simulations at different resolutions. The left panel compares the TPCF different  resolutions for TSF, while the right compares TPCFs for MSF. Each line represent the average TPCF for $150\lesssim t \lesssim 450\Myr$. The dotted vertical magenta line shows the mean radius of the GMCs found in TSF. To guide the reader the fits from Fig.~\ref{fig:stpcf} are included as the black dotted lines. 
		}
		\label{fig:restpcf}
	\end{center}
\end{figure}

Here we present the comparisons of the gas PDF, \Sfr and stellar TPCF for TSF and MSF when run at different maximum resolutions, Fig.~\ref{fig:resgpdf}, \ref{fig:rescm} and \ref{fig:restpcf} respectively. We note that the TPCF for TSF and MSF when using a maximum resolution $\Delta x\sim2.3\pc$ lacks the turn over at small $r_{\star}$ seen in all other resolutions. This is due to keeping $\delta r_{\star}=5\pc$ when calculating the TPCF for all resolutions and thus not resolving separation of particles less than approximately 2 cells widths for the high resolution simulations.

%----- FB Cumalitive Mass -----%
%----- ------- -----%

%----- FB Cumalitive Mass -----%
%----- ------- -----%

%----- FB Cumalitive Mass -----%
%----- ------- -----%

\end{document}